\documentclass[useAMS]{mn2e}
\usepackage{graphicx}

\title[Recurrent dust formation by WR 48a]
{Recurrent dust formation by WR 48a on a 30-year timescale\thanks{Based 
on observations collected at the European Southern Observatory, La Silla, Chile, 
the South African Astronomical Observatory, Sutherland, South Africa, and 
the Anglo Australian Telescope, Siding Spring, Australia.}}
\author[P. M. Williams et al.]
       {Peredur M. Williams$^1$\thanks{Email: pmw@roe.ac.uk}, Karel A. van der Hucht$^{2,3}$, 
        Francois van Wyk$^4$, \and Fred Marang$^4$, Patricia A. Whitelock$^{4,5}$, 
        Patrice Bouchet$^6$ and \and Diah Y. A. Setia Gunawan$^7$\\
   $^1$Institute for Astronomy,  
     University of Edinburgh, Royal Observatory, Edinburgh EH9 3HJ\\
   $^2$Space Research Organization Netherlands, Sorbonnelaan 2,
          NL-3584 CA Utrecht, The Netherlands\\
   $^3$Astronomical Institute `Anton Pannekoek', University of
          Amsterdam, Science Park 904, NL-1098 XH Amsterdam, The Netherlands \\
   $^4$South African Astronomical Observatory, P.O. Box 9, 7935 Observatory, 
       South Africa\\
   $^5$Astronomy, Cosmology and Gravity Centre, Astronomy Department, 
       University of Cape Town, 7700 Rondebosch, South Africa\\
   $^6$Service d'Astrophysique DSM/IRFU/SAp CEA - Saclay, L'Orme des Merisiers 
       b\^atiment 709, F-91191 Gif-sur-Yvette France \\
   $^7$Atacama Large Millimetre/submillimetre Array (ALMA), Alonso de Cordova 3107, 
       Vitacura, Santiago 763 0355, Chile
   }

\date{Accepted 2011 November 16.
      Received 2011 November 16 ;
      in original form 2011 August 2}

\pagerange{\pageref{firstpage}--\pageref{lastpage}}
\pubyear{2011}

\def\Lp{$L^{\prime}$}

\begin{document}

\maketitle

\label{firstpage}

\begin{abstract}
We present infrared photometry of the WC8 Wolf-Rayet system WR\,48a observed 
with telescopes at ESO, the SAAO and the AAT between 1982 and 2011 which show 
a slow decline in dust emission from the previously reported outburst in 1978--79 
until about 1997, when significant dust emission was still evident. This was 
followed by a slow rise, accelerating to reach and overtake the first (1978) 
photometry, demonstrating that the outburst observed in 1978--79 was not an 
isolated event, but that they recur at intervals of 32+ years. This suggests  
that WR\,48a is a long-period dust maker and colliding-wind binary (CWB). 
The locus of WR\,48a in the ($H-L$), $K$ colour-magnitude diagram implies 
that the rate of dust formation fell between 1979 and about 1997 and then 
increased steadily until 2011. 
Superimposed on the long-term variation are secondary (`mini') eruptions in 
(at least) 1990, 1994, 1997, 1999 and 2004, characteristic of relatively 
brief episodes of additional dust formation. 
Spectra show evidence for an Oe or Be companion to the WC8 star, supporting 
the suggestion that WR\,48a is a binary system and indicating a system 
luminosity consistent with the association of WR\,48a and the young star 
clusters Danks 1 and Danks 2. The range of dust formation suggests that 
these stars are in an elliptical orbit having $e \sim 0.6$. The size of the 
orbit implied by the minimum period, together with the WC wind velocity and 
likely mass-loss rate, implies that the post-shock WC wind is adiabatic 
throughout the orbit -- at odds with the observed dust formation. A similar 
conflict is observed in the `pinwheel' dust-maker WR\,112.
\end{abstract}


\newpage

\section{Introduction}

\noindent Population I Wolf-Rayet (WR) stars are characterised by dense, fast 
stellar winds giving the stars their broad emission-line spectra and carrying 
heavy mass loss: they represent the last stable stage in the evolution of massive 
stars. One of the earliest results from infrared (IR) astronomy was the 
discovery of `excess' radiation by heated circumstellar dust from a variety of 
stars known to be losing mass. Amongst these were three WC9 type WR stars, 
observed by Allen, Harvey \& Swings (1972). Subsequent studies (e.g. Cohen, 
Barlow \& Kuhi 1975) found circumstellar dust emission from most WC9 and 
many WC8 stars. The process was soon inverted, and bright IR sources discovered 
in space- or ground-based IR surveys were found to harbour optically faint dusty 
WC8--9 stars (e.g. AFGL\,2104 = WR\,112, Cohen \& Kuhi 1977).  Amongst these was  
WR\,48a, discovered by Danks et al. (1983, hereafter DDWS) in a survey of the 
H\,{\sc ii} region G305.5 +0.0. They observed WR\,48a to be undergoing an infrared 
outburst similar to that shown in 1977 by HD 193793 (Williams et al. 1978, 
Hackwell, Gehrz \& Grasdalen 1979). DDWS tracked the rise of the IR emission to 
maximum in 1979 and its initial fading, and presented a spectrum showing He and C 
emission lines, most conspicuously He\,{\sc i} 10830\AA\ and C\,{\sc iii} 9710\AA, 
from which they classified WR\,48a as a WC9 star. 

In their survey of IR dust emission by late subtype WC stars, Williams, 
van der Hucht \& Th\'e (1987a, hereafter WHT), reported photometry of WR\,48a 
in 1982--86 which showed continuing fading of its dust emission. They also 
extended the spectroscopy to the yellow (5550--5950\AA), and argued from the 
approximately equal intensities in the C\,{\sc iii} 5696-\AA\ and C\,{\sc iv} 
5808-\AA\ classification lines that WR\,48a was a WC8 star. Further IR photometry 
of WR\,48a was reported by Williams et al. (2003, hereafter Paper~I), showing 
that the fading continued to about 1997, but 
that the spectral energy distribution (SED) at that time still showed  dust 
emission, indicating that dust was still being formed by WR\,48a. 
In this respect it differed from `episodic' dust makers like HD 193793 (WR\,140), 
which form dust for only a fraction of their periods (Williams et al. 1990). 
The IR photometry of WR\,48a also showed secondary dust-formation episodes, 
`mini eruptions', in 1990 and 1994. These resembled those observed by Williams 
et al. (2001) in another episodic dust maker, the WC7+O9 system WR\,137 (HD 192641). 
In Paper~I, the 2.4--25-$\mu$m SED of WR\,48a observed with the short-wavelength 
spectrometer (SWS) on {\em ISO} in a survey of WR spectra (van der Hucht et al. 
1996) was compared with that of WR\,104 (Ve2--45), the classical dust-making 
WC9 star, and found to be significantly cooler. This difference was consistent 
with the radial dust density distribution around WR\,48a being flatter than 
the $\propto r^{-2}$ law expected of a system making dust at a constant rate 
like WR\,104: the evolution of the dust density distribution 
of WR\,48a is considered further in Section \ref{SModels}. The {\em ISO}-SWS 
spectrum of WR\,48a presented in Paper~I showed an emission feature near 
7.7 $\mu$m similar to that observed in WR\,104 and WR\,112 by Cohen, Tielens 
\& Bregman (1989) and ascribed to a carbonaceous carrier with small aromatic 
domains, a possible precurson to carbon grains; an independent study of the 
SWS spectrum by Chiar, Peeters \& Tielens (2002) identified and discussed 
emission features at 6.4$\mu$m and 7.9$\mu$m.

The enduring interest of dust formation by WR stars is the great difficulty 
in forming dust in such hostile environments: close to the stars, the stellar 
radiation fields would heat any dust to above its evaporation temperature 
but at greater distances, where the radiation field is sufficiently diluted, 
the density of the WR stellar wind is too low to allow homonuclear dust 
condensations -- as recognised early on by Hackwell et. al. (1979) in the case 
of WR\,140 and discussed by WHT for the WC8--9 stars making dust persistently. 
Zubko (1998) found that carbon 
grains having drift velocities relative to the WC8--9 wind could grow via 
collision with carbon ions, but the question of grain nucleation remains open. 
A chemical kinetic study of the formation of molecular species and carbon 
grain precursors in WC9 winds by Cherchneff et al. (2000) found that only 
C$_2$ was formed in useful amounts, and even that required significantly 
(factor $>10^3$) higher densities than expected in the regions of isotropic, 
smooth WR winds where dust grains are observed. 
Understanding the physicochemical process of dust formation by WR stars 
appears to have stalled, but the incidence of observed dust formation by 
WR stars in different systems provides valuable information on the origin 
of the significantly higher densities demanded. 

Large-scale, high-density structures in a WR wind are provided by compression of 
the stellar wind in strong shocks formed where the wind of a WR star collides 
with that of a sufficiently luminous companion in a binary system -- a colliding 
wind binary (CWB) -- as occurs in WR\,140. Usov (1991) suggested that very 
high compression factors ($10^{3} - 10^{4}$) could be produced in WR\,140 
if the heated and compressed wind was able to cool efficiently. 
The link between the dust-formation episodes and binary orbit in WR\,140 is 
provided by the periodic increases by factor of $\sim$ 40 of the pre-shock wind 
density at the wind-collision region (WCR) for a brief time around periastron 
passage (Williams 1999) in its very elliptical orbit. 
Similarly, Lef\`evre et al. (2005) showed that the episodic dust-maker WR\,137 
was a spectroscopic binary having a period (13.05 yr) equal to that of the 
dust-formation episodes (Williams et al. 2001), with periastron passage about 
a year before maximum dust emission. 
More recently, Williams, Rauw \& van der Hucht (2009a) showed that the radial 
velocities of the WC5+O9 system WR\,19 were consistent with its being a 
high-eccentricity ($e = 0.8$) binary, having periastron passage shortly 
before its brief, periodic (10.1-yr) dust-formation episodes.

The connection between the persistent dust-formation by the WC9 stars and their 
movement in binary orbits comes from IR images of the dust, which show rotating 
`pinwheel' structures around WR\,104, WR\,98a (IRAS 17380--3031) 
and WR\,112, interpreted to be dust formed and ejected in a stream to one side 
of a rotating binary system observed at a relatively low orbital inclination 
(Tuthill, Monnier \& Danchi 1999, Monnier, Tuthill, \& Danchi 1999, and 
Monnier et al. 2007).  
The observation of non-thermal radio emission from the `pinwheel' systems 
(Monnier et al. 2002) provides further support for their being CWBs; but their 
stellar components have not yet been resolved, so that it has not been 
possible to relate the positions of the stars and WCRs to that of the dust. 

Monnier et al. (2007) found the dust emission by WR\,48a to be extended at 
2$\mu$m; and the dust cloud was imaged at 12.3$\mu$m by Marchenko \& Moffat 
(2007), showing the dust emission to be extended E--W, suggesting a flattened 
structure, possibly a fragmentary spiral, seen at moderately high inclination.

DDWS and Danks et al. (1984) associated WR\,48a with two newly discovered 
galactic clusters, Danks~1 (C 1310--624) and Danks~2 (C 1309--624) at a 
distance between $\sim$ 3.3 and 4.6 kpc. Bica et al. (2004) derived 
colour-magnitude diagrams from deeper photometry and derived distances 
to the clusters of $3.6\pm0.5$ and $3.4\pm0.2$ kpc respectively. 
The connection between WR\,48a and the clusters was reinforced by 
Clark \& Porter (2004) in their study of star formation in the region. 
They suggested that star formation 
might have been triggered by WR\,48a and adopted a distance of 4 kpc 
for the complex. This gives a greater distance to WR\,48a than that 
(1.2~kpc, van der Hucht 2001) implied by its spectral type, and these 
estimates need to be reconciled.

By all accounts, WR\,48a is heavily reddened, but the exact value is not 
certain. DDWS derived A$_v \simeq$ 9.2 from the 7000--10000-\AA\ colour 
in their spectrum. Williams, van der Hucht \& Th\'e (1987b) derived 
A$_v \simeq 8.3$ from the 5000--7000-\AA\ slope in their spectrum observed 
in 1986, and noted that this was consistent with the strength of the 
6284-\AA\ diffuse interstellar absorption band (DIB) and the relation 
between these quantities from other reddened stars. 
From the $ISO$ SWS spectrum of WR\,48a, Schutte et al. (1998) found the 
interstellar `silicate' feature to have $\tau_{9.7} = 0.52$. The ratio of 
visual extinction to `silicate' optical depth in the spectra of WR stars 
and Cyg OB2 No.\ 12 showed significant scatter, $A_V/\tau_{9.7} = 16.6\pm4.6$.
Applying this ratio to the $\tau_{9.7}$ observed in WR\,48a gives 
$A_V = 8.6\pm2.4$. (A lower value, $A_V$ (6.1), was derived from $\tau_{9.7}$ 
in Paper~I but this used a different and probably inappropriate calibration.)
Baume, Carraro \& Momany (2009) observed $UBVIc$ photometry in 2006, and 
derived $E(B-V) \simeq 2.4$. For a standard reddening law, this implies 
a lower extinction and, from their observed $(V-Ic) = 3.88$, an intrinsic 
$(V-Ic) \simeq 0.9$, which is abnormally red for an early type star. 
If some of the extinction occurs in the carbon dust made by WR\,48a, the 
extinction law may not be standard. For example, Menzies \& Feast (1997) 
found an anomalous reddening law for the carbon dust made by the RCB star 
RY Sgr, so the sightline to WR\,48a may have a reddening law intermediate 
between the standard interstellar and that for carbon dust. 
On the other hand, the circumstellar component of the reddening is unlikely 
to be very great, given the strength of the `silicate' feature and the 
reddening to the associated clusters $A_V = 8.1$ and $9.3$ respectively 
(Bica et al.).  For the present, therefore, we adopt $A_V = 9.0$ and a 
standard interstellar law for the reddening of WR\,48a.

Zhekov, Gagn\'e \& Skinner (2011) have identified a strong, highly obscured 
X-ray source with WR\,48a. They pointed out that, if WR\,48a was physically 
associated with the clusters Danks 1 and Danks 2 at a distance of $\sim 4$kpc, 
it would be the most X-ray luminous WR star in the Galaxy, after Cyg X-3.
They also derived a high extinction, $N_H = 2.29 \times 10^{22}$ cm$^{-2}$, 
from their NEI shock modelling of the X-ray spectrum and noted that this was 
consistent with $A_V = 10$ using the standard conversion, 
$N_H = 2.22 \times 10^{21} A_V$ cm$^{-2}$; but $A_V$ could be 40-50\% higher 
using more recent conversions. As one of several possible origins for the 
X-ray emission, they considered colliding stellar winds in a wide WR+O binary.
As noted by Zhekov et al., and also Mauerhan, van Dyk \& Morris (2011), 
WR\,48a was detected with {\em Chandra} as a very strong X-ray source.

In this paper, we report almost three decades (1982--2011) of IR photometry 
tracking the emission by the dust cloud made by WR\,48a and showing that 
the 1978--9 dust-formation was not an isolated event, but one which appears 
to be recurring after an interval of over 32 yrs. 
We present optical spectra which suggest that the WC8 star in WR\,48a has an 
OB companion, apparently an Oe or Be star. 
We use the IR colours and simple modelling of the SEDs to describe the 
evolution of the dust cloud and develop a preliminary picture of WR\,48a 
as a colliding-wind binary.

\section{Observations}\label{SObs}      

\subsection{Infrared photometry}

\begin{table}
\caption{Near-infrared photometry of WR\,48a with ESO telescopes: except 
where marked (* 3.6m, + 2.2m), observations were made with the 1-m 
photometric telescope.}
\label{TESOa} 
\centering
\begin{tabular}{llllll}        
\hline
Date    &  $J$ &  $H$ &  $K$ &  \Lp &  $M$  \\ 
\hline
2445418 & 7.80 & 5.21 & 3.01 & 0.29 & -0.42 \\ 
2445419 & 7.88 & 5.25 & 3.10 & 0.41 & -0.37 \\ 
2445839 & 8.27 & 5.61 & 3.39 & 0.52 & -0.31 \\ 
2446278 & 8.46 & 5.94 & 3.71 & 0.80 &  0.04 \\ %
2446632 & 8.57 & 6.23 & 4.07 & 1.13 &  0.27 \\ %
2446829 & 8.68 & 6.40 & 4.22 & 1.18 &  0.32 \\ %
2446831 & 8.69 & 6.38 & 4.12 & 1.27 &  0.25 \\ %
2447258 & 8.72 & 6.46 & 4.35 & 1.41 &  0.58 \\ 
2447259 & 8.73 & 6.45 & 4.32 & 1.36 &  0.63 \\ 
2447260 & 8.70 & 6.45 & 4.33 & 1.40 &  0.63 \\ 
2447261 & 8.70 & 6.46 & 4.36 & 1.42 &  0.61 \\  
2447519 & 8.91 & 6.69 & 4.57 & 1.54 &  0.64 \\ 
2447520 & 8.94 & 6.69 & 4.57 & 1.54 &  0.73 \\ 
2447540 & 9.17 & 6.78 & 4.56 & 1.59 &  0.67 \\  
2447453 & 8.94 & 6.75 & 4.55 & 1.56 &  0.66 \\ 
2447544 & 8.93 & 6.70 & 4.56 & 1.58 &  0.69 \\ 
2447552 & 8.85 & 6.73 & 4.55 & 1.55 &  0.70 \\ 
2447573 & 8.87 & 6.71 & 4.60 & 1.58 &  0.74 \\ 
2447670 & 8.73 & 6.67 & 4.63 & 1.69 &  0.91 \\  
2447671 & 8.86 & 6.80 & 4.70 & 1.66 &  0.73 \\ 
2447672 & 8.86 & 6.82 & 4.73 & 1.71 &  0.78 \\ 
2447910 & 8.89 & 6.83 & 4.80 & 1.79 &  0.89 \\ 
2447966 & 8.87 & 6.87 & 4.86 & 1.85 &  0.94 \\ 
2447967 & 8.93 & 6.89 & 4.86 & 1.87 &  0.96 \\ 
2447968 & 8.90 & 6.87 & 4.83 & 1.83 &  0.92 \\ 
2447970 & 8.91 & 6.88 & 4.85 & 1.84 &  0.96 \\ 
2447971 & 8.90 & 6.88 & 4.84 & 1.83 &  0.95 \\ 
2447991 & 8.91 & 6.89 & 4.87 & 1.87 &  0.93 \\ 
2447992 & 8.86 & 6.88 & 4.86 & 1.84 &  0.95 \\ 
2447993 & 8.89 & 6.88 & 4.87 & 1.85 &  0.97 \\ 
2448316 & 8.89 & 6.91 & 4.89 & 1.94 &  1.01: \\ 
2448317 & 8.89 & 6.89 & 4.89 & 1.92 &  1.02 \\ 
2448318 & 8.91 & 6.90 & 4.88 & 1.93 &  1.02 \\ 
2448347* & 8.91 & 6.93 & 4.93 & 1.97 &  1.07\\ 
2448348 & 8.94 & 6.96 & 4.95 & 1.97 &  1.04 \\ 
2448349 & 8.96 & 6.96 & 4.96 & 1.98 &  1.09 \\ 
2448350 & 8.95 & 6.98 & 4.97 & 1.97 &  1.08 \\ 
2448639 & 8.91 & 6.92 & 4.95 & 2.03 &  1.16 \\
2448698 & 8.96 & 7.02 & 5.04 & 2.08 &  1.14 \\ 
2448749 & 8.98 & 7.07 & 5.10 & 2.11 &  1.21 \\ 
2448752 & 9.00 & 7.07 & 5.10 & 2.12 &  1.20 \\ 
2448753 & 8.92 & 7.05 & 5.09 & 2.12 &  1.21 \\ 
2448799 & 9.16 & 7.14 & 5.13 & 2.13 &  1.20 \\  
2449050+ & 9.01 & 7.15 & 5.24 & 2.26 &  1.35 \\ 
2449052+ & 8.98 & 7.14 & 5.24 & 2.26 &  1.35 \\ 
2449179 & 8.97 & 7.09 & 5.20 & 2.29 &       \\ 
2449180 & 8.96 & 7.10 & 5.21 & 2.30 &  1.37 \\  
2449181 & 8.94 & 7.11 & 5.22 & 2.31 &  1.38 \\ 
2449182 & 8.92 & 7.11 & 5.23 & 2.32 &  1.39 \\  
2449183 & 8.96 & 7.11 & 5.22 & 2.31 &  1.36 \\  
2450198+ &      & 7.11 & 5.37 & 2.49 &  1.61 \\ 
2450890+ & 9.07 & 7.12 & 5.20 & 2.54 &  1.78 \\   
\hline
\end{tabular}
\end{table}

\begin{table}
\caption{Long-wavelength photometry of WR\,48a with ESO telescopes}
\label{TESOb} 
\begin{tabular}{rlrlllll}        
\hline
Date    &  \Lp &  $M$  & $N1$ &  $N2$ &  $N3$  &  $Q0$ & Tel. \\ 
\hline
2445150 & 0.08 & -0.67 & -1.6 & -1.4  &  -1.9  &  -1.7 & 3.6m \\ 
2445425 & 0.29 & -0.46 & -1.4 & -1.3  &  -1.6  &  -1.5 & 3.6m \\
2446194 & 0.86 & -0.03 & -1.3 & -1.1  &  -1.7  &       & 3.6m \\ 
2447519 & 1.54 &  0.73 & -0.9 & -0.5  &  -1.4  &       & 1.0m \\ %
2447987 & 1.81 &  0.80 & -0.8 & -0.6  &  -1.2  & -1.2  & 2.2m \\ 
2447988 & 1.88 &  0.91 & -0.7 & -0.5  &  -1.3  & -1.3  & 2.2m \\ 
2448346 & 1.95 &  1.10 & -0.6 & -0.4  &  -1.1  & -1.2  & 3.6m \\ 
2448347 & 1.96 &  1.09 & -0.6 & -0.3  &  -1.1  & -1.1  & 3.6m \\ 
\hline
\end{tabular}
\end{table}

\begin{table*}
\caption{SAAO photometry of WR\,48a with the Mk II IR photometer on the 0.75-m 
telesope using the 36\arcsec (*or 24\arcsec) aperture.}
\label{TSAAO}
\begin{minipage}{5.5cm}
\centering
\begin{tabular}{lcccc}
\hline
Date & $J$ & $H$  &  $K$  &  $L$  \\
\hline
2445417 & 7.78 & 5.19 & 3.05 & 0.65 \\ 
2447547 & 8.38 & 6.50 & 4.50 & 1.98 \\ 
2447553 & 8.34 & 6.49 & 4.57 & 1.98 \\ 
2447607 & 8.34 & 6.55 & 4.57 & 2.00 \\ 
2447634 & 8.42 & 6.64 & 4.63 & 2.11 \\ 
2447669 & 8.39 & 6.59 & 4.62 & 2.08 \\ 
2447687 & 8.43 & 6.63 & 4.63 & 2.08 \\ 
2447737 & 8.42 & 6.65 & 4.67 & 2.17 \\ 
2448028 & 8.36 & 6.66 & 4.77 & 2.24 \\ 
2448108 & 8.31 & 6.46 & 4.61 & 2.30 \\ 
2448285 & 8.40 & 6.87 & 4.78 & 2.29 \\ 
2448289 & 8.42 & 6.70 & 4.78 & 2.31 \\ 
2448327 & 8.41 & 6.73 & 4.82 & 2.34 \\ 
2448365 & 8.43 & 6.76 & 4.87 & 2.35 \\ 
2448378 & 8.38 & 6.73 & 4.85 & 2.41 \\ 
2448385 & 8.41 & 6.76 & 4.87 & 2.35 \\ 
2448427 & 8.42 & 6.74 & 4.88 & 2.42 \\ 
2448437 & 8.41 & 6.74 & 4.89 & 2.41 \\ 
2448451 & 8.40 & 6.73 & 4.86 & 2.39 \\ 
2448457 & 8.41 & 6.71 & 4.85 & 2.39 \\ 
2448462 & 8.40 & 6.72 & 4.85 & 2.38 \\ 
2448500 & 8.33 & 6.62 & 4.78 & 2.37 \\ 
2448701*& 8.47 & 6.80 & 4.96 & 2.48 \\ 
2448734*& 8.54 & 6.83 & 4.98 & 2.51 \\ 
2448760*& 8.44 & 6.83 & 5.00 & 2.51 \\ 
2448783 & 8.40 & 6.82 & 5.00 & 2.52 \\ 
2448792 & 8.46 & 6.85 & 5.01 & 2.54 \\ 
2448817 & 8.45 & 6.86 & 5.03 & 2.56 \\ 
2448851 & 8.51 & 6.90 & 5.07 & 2.63 \\ 
2449001 & 8.48 & 6.94 & 5.11 & 2.60 \\ 
2449106 & 8.52 & 6.91 & 5.14 & 2.68 \\ 
2449146 & 8.42 & 6.87 & 5.12 & 2.69 \\ 
2449221 & 8.44 & 6.90 & 5.14 & 2.75 \\ 
\hline
\end{tabular}
\end{minipage}
\hfill
\begin{minipage}{5.5cm}
\centering
\begin{tabular}{lcccc}
\hline
 Date & $J$ & $H$  &  $K$  &  $L$  \\
\hline
2449382 & 8.56 & 6.97 & 5.23 & 2.79 \\  
2449406 & 8.46 & 6.86 & 5.14 & 2.77 \\  
2449501 & 8.35 & 6.62 & 4.91 & 2.68 \\  
2449519 & 8.28 & 6.55 & 4.85 & 2.67 \\  
2449583 & 8.32 & 6.58 & 4.84 & 2.62 \\  
2449590 & 8.34 & 6.59 & 4.84 & 2.59 \\  
2449735 & 8.43 & 6.75 & 4.96 & 2.61 \\  
2449820 & 8.46 & 6.79 & 4.98 & 2.68 \\  
2449887 & 8.46 & 6.85 & 5.04 & 2.68 \\  
2450091 & 8.47 & 6.91 & 5.13 & 2.76 \\  
2450127 & 8.44 & 6.89 & 5.13 & 2.77 \\  
2450178 & 8.45 & 6.95 & 5.21 & 2.75 \\  
2450206 & 8.42 & 6.95 & 5.20 & 2.83 \\  
2450237 & 8.47 & 6.95 & 5.22 & 2.86 \\  
2450259 & 8.49 & 6.95 & 5.23 & 2.90 \\  
2450320 & 8.48 & 6.99 & 5.28 & 2.90 \\  
2450484 & 8.46 & 7.02 & 5.32 & 2.90 \\  
2450591 & 8.46 & 7.07 & 5.44 & 3.03 \\  
2450618 & 8.52 & 7.06 & 5.43 & 3.07 \\  
2450798 & 8.41 & 6.82 & 5.12 & 2.89 \\  
2450980 & 8.44 & 6.92 & 5.23 & 2.98 \\  
2451234 & 8.40 & 6.83 & 5.14 & 2.92 \\  
2451298 & 8.46 & 6.93 & 5.23 & 2.94 \\  
2451358 & 8.45 & 6.93 & 5.21 & 2.99 \\  
2451576 & 8.34 & 6.68 & 5.02 & 2.89 \\  
2451614 & 8.39 & 6.71 & 4.99 & 2.86 \\  
2451677 & 8.42 & 6.75 & 5.02 & 2.86 \\  
2451685 & 8.45 & 6.77 & 5.02 & 2.87 \\  
2451715 & 8.42 & 6.79 & 5.05 & 2.84 \\  
2451743 & 8.43 & 6.82 & 5.06 & 2.86 \\  
2451785 & 8.44 & 6.84 & 5.08 & 2.88 \\  
2451967 & 8.42 & 6.86 & 5.14 & 2.93 \\  
\hline                                  
\end{tabular}                           
\end{minipage}
\hfill
\begin{minipage}{5.5cm}
\centering
\begin{tabular}{lcccc}
\hline
Date & $J$ & $H$  &  $K$  &  $L$  \\
\hline
2451981 & 8.42 & 6.86 & 5.14 & 2.89 \\   
2452070 & 8.40 & 6.78 & 5.06 & 2.90 \\   
2452093 & 8.41 & 6.79 & 5.06 & 2.87 \\   
2452116 & 8.45 & 6.80 & 5.07 & 2.88 \\   
2452290 & 8.52 & 6.85 & 5.10 & 2.90 \\   
2452322 & 8.41 & 6.81 & 5.08 & 2.91 \\   
2452351 & 8.41 & 6.84 & 5.12 & 2.90 \\ 
2452383 & 8.42 & 6.80 & 5.11 & 2.94 \\   
2452426 & 8.36 & 6.75 & 5.07 & 2.93 \\   
2452451 & 8.38 & 6.74 & 5.05 & 2.91 \\   
2452694 & 8.44 & 6.87 & 5.18 & 2.97 \\   
2452755 & 8.36 & 6.72 & 5.03 & 2.92 \\   
2452798 & 8.55 & 6.82 & 5.05 & 2.94 \\   
2452868 & 8.38 & 6.68 & 4.97 & 2.86 \\   
2453065 & 8.41 & 6.72 & 4.92 & 2.76 \\   
2453118 & 8.36 & 6.66 & 4.90 & 2.74 \\   
2453177 & 8.22 & 6.42 & 4.70 & 2.61 \\   
2453194 & 8.24 & 6.39 & 4.66 & 2.60 \\   
2453408 & 8.26 & 6.37 & 4.54 & 2.43 \\   
2453442 & 8.31 & 6.42 & 4.55 & 2.42 \\   
2453507 & 8.30 & 6.43 & 4.60 & 2.43 \\   
2453562 & 8.28 & 6.38 & 4.55 & 2.45 \\   
2453906 & 8.19 & 6.05 & 4.19 & 2.14 \\   
2453935 & 8.13 & 6.06 & 4.20 & 2.13 \\   
2454184 & 7.99 & 5.77 & 3.88 & 1.84 \\ 
2454234 & 7.93 & 5.70 & 3.81 & 1.75 \\   
2455011 & 7.52 & 5.04 & 3.16 & 1.08 \\   
2455353 & 7.56 & 4.90 & 2.99 & 0.99 \\   
2455621 & 7.21 & 4.71 & 2.86 & 0.82 \\   
2455642 & 7.25 & 4.75 & 2.92 & 0.99 \\   
2455731 & 7.26 & 4.73 & 2.84 & 0.77 \\   
2455733 & 7.24 & 4.72 & 2.85 & 0.77 \\   
\hline                                   
\end{tabular}
\end{minipage}
\end{table*}

Photometry of WR\,48a was obtained using aperture photometers on the 1-m, 2.2-m 
and 3.6-m telescopes of the European Southern Observatory (ESO) at La Silla, Chile, 
(Tables \ref{TESOa} and \ref{TESOb}), the Mk II photometer on 0.75-m telescope of 
the South African Astronomical Observatory (SAAO), Sutherland, South Africa 
(Table \ref{TSAAO}), and the Infrared Photometer/Spectrometer (IRPS) on the 
Anglo-Australian Telescope (AAT) at Siding Spring, Australia (Table \ref{TAAT}). 
The final two observations in Table \ref{TESOa} were obtained with the array camera 
IRAC-1 on the 2.2-m telescope at La Silla.

During our campaign, the near-infrared (NIR) photometric systems on the three ESO 
telescopes were the same, and as described by Bouchet, Manfroid \& Schmider (1991) 
and Bersanelli, Bouchet \& Falomo (1991). The final (`Postscript') photometry by 
DDWS was on this system, whereas that in the body of their paper was on the earlier 
ESO system (Wamsteker 1981), which differed principally in using a filter having 
$\lambda_{\rmn{eff}} \simeq$ 3.5--3.6 $\mu$m instead of 3.8 $\mu$m for $L$. 
The 3.8-$\mu$m filter is often called `\Lp' to distinguish it from the 
shorter-wavelength $L$, a convention we follow here. 
The SAAO photometry (Carter 1990) uses a shorter-wavelength ($\lambda_{\rmn{eff}} 
\simeq 3.5 \mu$m) $L$ filter. 
The  NIR system at the AAT (Allen \& Cragg 1983) also used a 3.8-$\mu$m filter 
for \Lp, but their $J$ filter had a slightly shorter $\lambda_{\rmn{eff}}$ than 
the ESO $J$ filter and closer to that of the $J$ filter in the SAAO Mk II 
photometer (see Glass 1993). 

Observations of WR\,48a can be affected by the nearby 11th magnitude star 
CPD -62$\degr$ 3058. From HST WFPC2 images, Wallace et al. (2003) measured a 
separation $9\farcs2$ and position angle $194\fdg2$, placing it within or 
near the edges of some of the apertures used for the photometry of WR\,48a. 
To determine the effect of this contribution on the photometry, we observed 
CPD -62$\degr$ 3058 with IRIS on the AAT through a 5$\arcsec$ aperture and give the 
results, together with magnitudes from the DENIS (Epchtein et al. 1999) and 2MASS 
(Skrutskie et al. 2006) surveys, in Table \ref{TCPD}. Despite the scatter in 
the data, we assume that the CPD star is not significantly variable in the NIR. 
In the $J$ band, the CPD star is 0.2 mag.\, fainter than WR\,48a at its minimum 
(2.4 mag. at maximum) and so makes a significant contribution to the $J$ 
magnitudes observed with the SAAO 0.75-m telescope, which used 36$\arcsec$ or, 
occasionally, 24$\arcsec$ apertures (see Table \ref{TSAAO}). 
This contribution is much smaller at longer wavelengths, with differences (WR\,48a 
minus CPD star) in the ranges $\Delta H = 2 - 5$ mag., $\Delta K = 3.5-6.2$ mag., 
and $\Delta L \simeq 5-8$ mag. corresponding to minima and maxima in WR\,48a. 
The AAT photometry used apertures of 5$\arcsec$ or 6$\arcsec$ and that with the 
ESO 3.6-m telescope used a $7\farcs5$ aperture, so the CPD star should have been 
well excluded.
Depending on the seeing and centring, there might have been some contamination 
of observations with the ESO 1-m and 2.2-m photometers, which generally used 
15$\arcsec$ and 12$\arcsec$ apertures. We searched for this by comparing the 
dispersion in the $J$ magnitudes when WR\,48a was observed on successive nights 
in an observing run (Table \ref{TESOa}) with the dispersion in $K$, 
where any contamination by the CPD star would be at least a factor of 20 smaller, 
but found no significantly greater dispersion in $J$.

\begin{table}
\caption{Infrared photometry of WR\,48a with the IRPS on the AAT}
\label{TAAT}
\centering
\begin{tabular}{lccccc}
\hline
Date    & $J$  &  $H$ &  $K$ &  \Lp & $M$   \\
\hline
2446309 & 8.61 & 5.94 & 3.73 & 1.00 & -0.11 \\ 
2446595 & 8.88 & 6.17 & 4.00 & 1.11 &       \\  
2446716 & 8.98 & 6.28 & 4.07 & 1.18 &       \\  
2446888 & 8.76 & 6.40 & 4.26 & 1.31 &       \\  
2447614 & 8.97 & 6.77 & 4.70 & 1.71 &       \\  
\hline
\end{tabular}
\end{table}

\begin{table}
\caption{Infrared photometry of CPD -62$\degr$ 3058 with the IRPS or from surveys.}
\label{TCPD}
\centering
\begin{tabular}{lcccccc}
\hline
Date    & $J$  &  $H$ & $Ks$ &  $K$ &  \Lp & source  \\
\hline
2446888 & 9.48 & 9.20 &      & 8.99 &      & AAT  \\ 
2447614 & 9.39 & 9.13 &      & 8.97 & 8.53 & AAT  \\
2450501 & 9.33 &      & 8.86 &      &      & DENIS \\ 
2450973 & 9.28 &      &      &      &      & DENIS \\ 
2451595 & 9.30 & 9.13 & 8.90 &      &      & 2MASS \\ 
\hline
\end{tabular}
\end{table}

\subsection{Homogenizing the NIR photometry}

\begin{table}
\caption{Zero points for the photometric systems employed and our {\em ad hoc} 
adjustments applied to the $L/L^{\prime}$ photometry of WR\,48a, the bands being 
identified by their wavelengths.}
\label{Tzero}
\centering
\begin{tabular}{lcccccc}
\hline
System          &  $H$  &  $K$  &  3.5  &  3.6  &  3.8  \\
\hline
AAT             & -0.02 & -0.01 &       &       & -0.03 \\
pre-1982 ESO    & -0.03 & -0.01 &       & 0.04  &       \\
post-1982 ESO   & -0.06 & -0.03 &       &       & -0.01 \\
SAAO            & 0.00  & 0.00  &  0.00 &       &       \\
adjustments  &       &       &  0.00 & 0.18  & 0.38  \\
\hline
\end{tabular}
\end{table}    
  
In order to determine the evolution of the dust emission by WR\,48a from the IR 
photometry, it is necessary to consider the relations between the different systems. 
The zero points of the systems are fairly similar (Table \ref{Tzero}, compiled from 
Cohen et al. 1999) but the redness of WR\,48a compared with the photometric standards 
is more of a problem.
Bouchet et al. (1991), 
Carter (1990), and Allen \& Cragg (1983) all derive colour transforms between their 
and other NIR photometric systems, none of these extend to objects as red as 
WR\,48a: most comparisons are based on stars having ($J-K$) $<$ 1.1, whereas 
WR\,48a generally has ($J-K$) $\simeq 4.0$. 
We therefore estimated {\em ad hoc} transformations for WR\,48a, concentrating 
on the various $L$ and $L^{\prime}$ systems, as the filters for these have very 
different wavelengths. We relate the AAT and two sets of ESO photometry 
to the corrected SAAO data as the last forms the largest homogeneous dataset. 

To relate the post-1982 ESO photometry of WR\,48a to the corrected SAAO photometry, 
we compared contemporaneous observations, i.e., those made within about 0.02 yr of 
each other, sometimes using successive observations at one observatory to bracket 
those made at the other and avoiding times when WR\,48a appeared to be changing rapidly. 
This gave eight useful comparisons between 1983 and 1993, yielding differences in 
the sense SAAO minus ESO: $\Delta J = -0.09\pm0.04$, $\Delta H = 0.07\pm0.03$, 
$\Delta K = 0.04\pm0.01$ and $L-L^{\prime} = 0.38\pm0.06$, while the ESO ($J-K$) 
colour of WR\,48a was in the range 4.8--3.9. The difference $L-L^{\prime}$ is 
much greater than that found by Bouchet et al. (1991) for stars having 
($J-K$)$<$1.0, but not surprising given the difference between the two 
filters and the extreme redness of WR\,48a.  
Attempts to derive a colour equation, such as a linear fit of the differences to 
($J-K$), did not yield statistically significant coefficients, so we restricted 
ourselves to applying a constant difference, $L-L^{\prime} = 0.38$, to our ESO 
photometry (and to the final, 1982, observation by DDWS) of WR\,48a to bring the 
data into line with the SAAO photometry for the production of homogeneous light 
and colour curves.  

For the earlier ESO photometry of WR\,48a in DDWS, we do not have 
contemporaneous SAAO observations for direct comparison. 
Instead, we used the photometry of WR stars by Pitault et al. (1983), which was also 
on the older ESO system and included two dust-emitting WC9 stars having ($J-K$) 
comparable to that of WR\,48a as an intermediate step. Comparison of their 
photometry of WR\,104 and WR\,106 with observations of these stars by WHT, which 
was on the post-1982 ESO system, suggests an offset $L$--\Lp = 0.20. 
This is about half the difference between \Lp and the SAAO $L$, 
as one might expect from the fact that the efective wavelength of the older ESO $L$ 
filter lies between those of \Lp and the SAAO $L$. To convert DDWS $L$ magnitudes of 
WR\,48a to the SAAO $L$ scale, we therefore added 0.18 mag., and made no adjustments 
to $H$ or $K$.

Four of the AAT observations of WR\,48a are contemporaneous with observations at ESO 
or SAAO, from which we infer an offset $L$--\Lp = 0.34, comparable to that for the 
ESO photometry, as one might expect from the similarity of the filter wavelengths.
Inspection of the AAT $H$ and $K$ data suggests that there is no significant shift 
between them and the ESO data.  

We applied these differences to the AAT and the two sets of ESO photometry and 
combined them with the corrected SAAO photometry to produce a homogeneous photometry 
set in $HKL$. The light curves from these data and the mid-IR photometry in Table 
\ref{TESOb} are discussed below, and the trajectory in the near-IR colour-magnitude 
diagram is discussed in Section \ref{SCMD}. The adjusted $L/L^{\prime}$ magnitudes 
are not used for production of the SEDs in Sections \ref{SSED} and \ref{SModels} 
where the calibrated fluxes are entered at their respective wavelengths.

\subsection{Optical spectroscopy}
\label{SOpt}

\begin{figure}
\centering
\includegraphics[width=9cm]{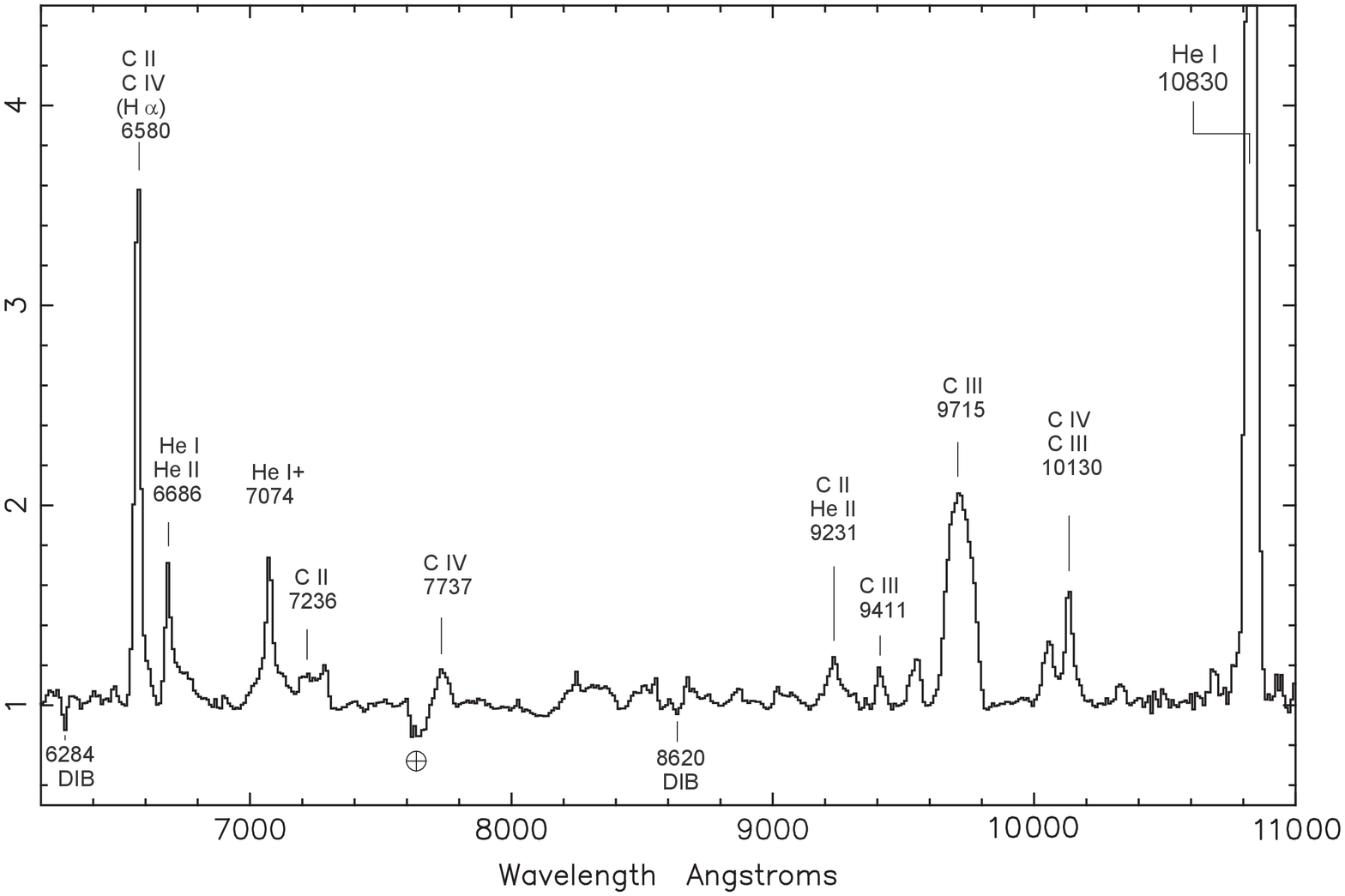}
\includegraphics[width=8.8cm]{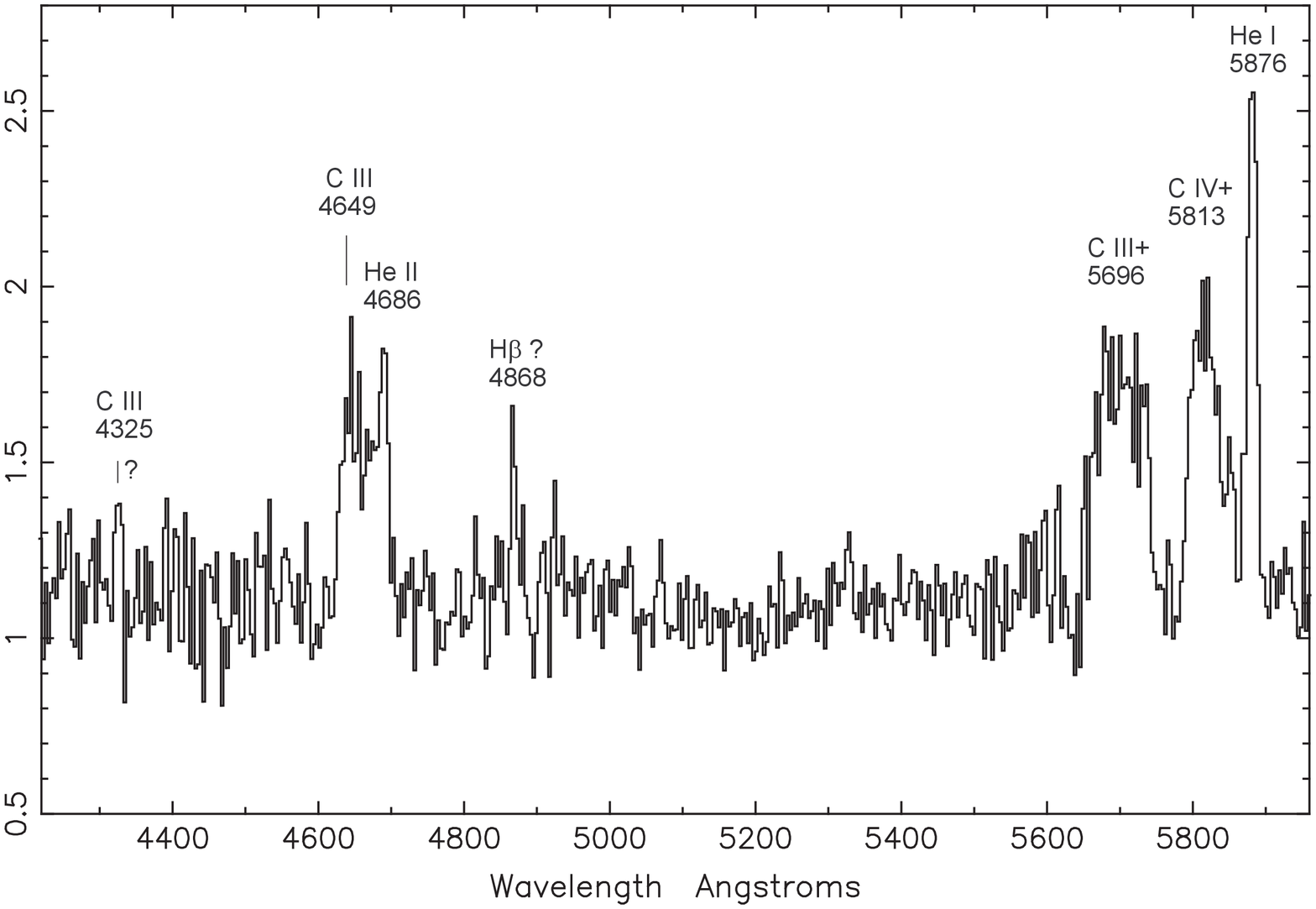}   
\includegraphics[width=8.5cm]{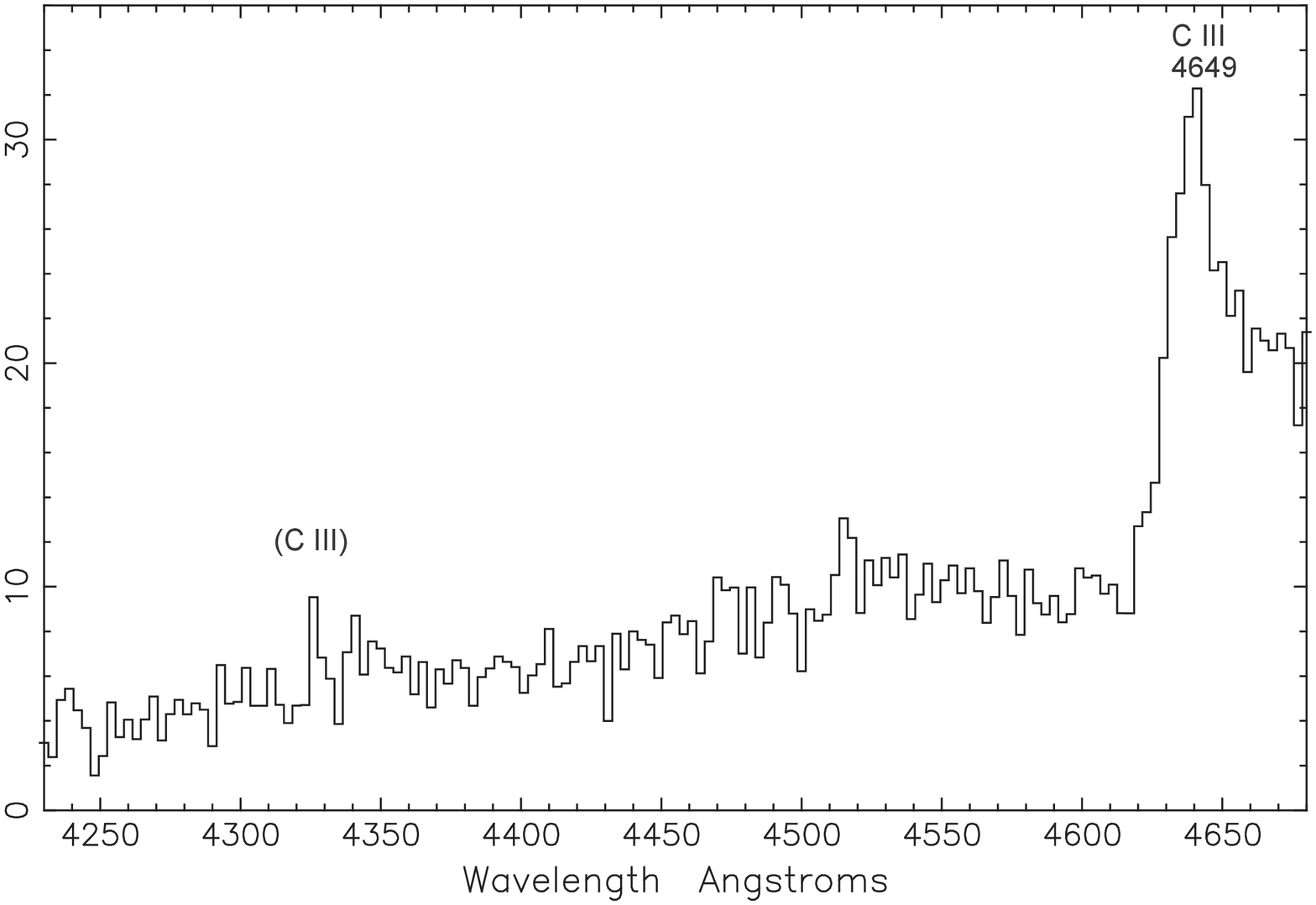}
\caption{Optical spectra of WR\,48a. Top: 18-\AA-resolution spectrum observed in 
1993. The 10830-\AA\ line is truncated to allow the weaker lines to be more easily 
seen. Middle: 7.4-\AA-resolution spectrum observed with the blue arm on the same time.
Bottom: blue spectrum of WR\,48a observed in 2001 with EMMI on the NTT, 
blocked to 3-\AA\ bins.}
\label{FOpt}
\end{figure}

New spectra were observed for us with the RGO spectrograph on the Anglo 
Australian Telescope on 1993 June 21, when the NIR emission had faded about 
2.5 mag.\ from maximum (cf. Fig\,\ref{FHKL}). In the blue, the resolution 
was 7.4\AA\ while that in the red, observed simultaneously, was 18\AA. 
A spectrum of the metal-poor star HD 184711 observed on the same night was 
used to help with removal of telluric absorption features in the red. 
Rectified spectra are shown in the first two panels of Fig. \ref{FOpt}. 
The correction for the strong telluric absorption in the A band (7630\AA) 
was incomplete because HD 184711 was observed at a rather lower airmass 
than WR\,48a. 

The `red' spectrum resembles that in DDWS, but the emission lines appear 
more strongly. 
We measured equivalent widths (EWs) of the strong He\,{\sc i} 10830-\AA\ 
and C\,{\sc iii} 9715-\AA\ lines to be 280\AA\ and 118\AA\ respectively, 
compared with EWs of $\sim$ 63\AA\ and 46\AA\ estimated from fig.\,3 of DDWS. 
Taken at face value, these differences suggest that, at the time (1980 January) 
DDWS's spectrum was observed, these emission lines were diluted by additional 
contimuum emission of factors 1.6 -- 3.4 above the stellar continuum between 
9715\AA\ and 10830\AA\, possibly by the newly formed dust, which is considered 
below. 

The 10830-\AA\ emission line falls close to the blue edge of the $J$ filter 
and line emission can contribute to the $J$ magnitudes. For example, the 
transmission of the $J$ filter in the SAAO Mk II IR photometer (Glass 1993)  
is about 50\% maximum at this wavelength while that of the post-1982 ESO $J$ 
filter (Bersanelli et al.) is nearer 20\%. The other optical components and 
the atmosphere will reduce these percentages. 
The corresponding contributions of the 10830-\AA\ line to $J$ magnitudes 
observed with the Mk II photometer at the times of the 1980 and 1993 spectra 
would then be 0.01 and 0.05 mag. for a flat continuum across the $J$-filter 
passband, and somewhat less for a rising dust continuum, while the contributions 
to the ESO $J$ magnitudes would be about half these values. We did not adjust 
$J$ photometry {\em en masse} but did adjust the SAAO $J$ magnitude used for 
the SED near minimum (Section \ref{SSED}) for the 10830-\AA\ line contribution.

We were not able to measure a P~Cygni absorption component of the 10830-\AA\ 
line, but its FWZI of 90~\AA\ allows us to estimate the wind velocity to be 
$v_{\infty} \simeq 1200\pm170$ km~s$^{-1}$. This is lower than the average 
(1700 km~s$^{-2}$) for WC8 stars (van der Hucht 2001, table 27) and equal to 
that for WC9 stars. This line needs to be re-observed  at higher resolution 
and higher S/N ratio to to refine our measurement. (The broad C\,{\sc iii} 
feature at 9715\AA\ is a multiplet with components ranging 9704--9720\AA.)

We measured the 6284-\AA\ diffuse interstellar band to have an EW of 3.1~\AA, 
comparable to the value (2.8\AA) in the 1986 spectrum, observed when the dust 
emission was greater (Williams, van der Hucht \& Th\'e 1987b) , so it is 
unlikely that this feature has a circumstellar component.

After the 10830-\AA\ line, the strongest emission line is that at 6570\AA\ 
(EW 73\AA). This is much stronger than the C\,{\sc iii} 9715-\AA\ feature, 
whereas it is weaker than the C\,{\sc iii} lines in the spectra of the WC8+O9 
systems WR\,11 ($\gamma$ Vel) and WR\,70 presented by Vreux, Dennefeld \& 
Andrillat (1983), and significantly weaker in the single WC8 star in their 
sample, WR\,53 (since reclassified WC9 by Crowther, de Marco \& Barlow 1998). 
The identification of the 6570-\AA\ feature is difficult. 
DDWS identified it with $\lambda$6560 He\,{\sc ii} (Hu$\beta$), 
but we consider any contribution of He\,{\sc ii} -- which we cannot rule 
out given the resolution of the spectrum -- to be small, because of the 
weakness or absence of the Hu$\alpha$ line at 10124\AA. There may be a 
small Hu$\alpha$ contribution to the C\,{\sc iii-iv} feature at 10130\AA, 
but the Hu$\beta$ line would be even smaller. 
Another contributor to the 6570\AA\ feature is multiplet~2 of C\,{\sc ii}.
The spectra of WR\,11, WR\,53 and WR\,70 (Vreux et al.) suggest that the 
strength of this should be comparable (within a factor of 2) to that of 
multiplet~3 of C\,{\sc ii} at 7236\AA; but the 6570-\AA\ line is a factor 
of about 16x stronger in our spectrum of WR\,48a, so C\,{\sc ii} cannot be 
the dominant contributor. Another possible 
contributor to the 6570\AA\ feature in low-resolution spectra of WR\,48a is 
the C\,{\sc iv} line at 6592\AA. This line accounts for the rising ratio of 
the 6570\AA\ and 7236\AA\ C\,{\sc ii} lines in the spectra of earlier 
subtype WC stars (cf. Vreux et al.), but nowhere is the ratio as high as it 
is observed in WR\,48a -- nor are the other C\,{\sc iv} lines in our 
spectrum found to be anomalously strong. 
No other plausible WC8 lines seem capable of explaining 
the strength of the 6570\AA\ feature. We suggest it comes from H$\alpha$, 
either from a nebula very close to the star (not resolved in the long-slit 
spectrum) or from an Oe or Be stellar companion to the WC star. 

Owing to the heavy reddening of WR\,48a, the blue spectrum 
is very noisy at short wavelengths, and only that longward of 4200\AA\ is 
considered useful. The only convincing feature in the blue is the blend of 
C\,{\sc iii} 4649\AA\ and He\,{\sc ii} 4686\AA. This is very weak, having an 
equivalent width (EW) of 36\AA, compared with the EW of 410\AA\ of the same 
feature measured from the low-resolution spectrum of the single WC8 star 
WR\,135 observed by Torres \& Massey (1987). There is a possible feature at 
the position of the 4325-\AA\ C\,{\sc iii} line and a stronger, narrow line 
at 4868 \AA\ which might be H$\beta$ emission from the proposed companion.

The spectrum in the yellow is similar to that observed in 1986 (WHT). 
Compared with the WC8 spectra in Torres \& Massey, the He\,{\sc i} 5896-\AA\ 
line in WR\,48a appears abnormally strong relative to the C\,{\sc iii} and 
C\,{\sc iv} lines. This is not a question of the classification of WR\,48a: 
the WC9 spectra in Torres \& Massey have the He\,{\sc i} comparable in 
strength to the C\,{\sc iv} line but much weaker than C\,{\sc iii}.
The carbon lines are much weaker: the EWs of the 5696-\AA\ and 
5813-\AA\ features in WR\,48a are 49~\AA\ and 41~\AA\ repectively, one 
quarter their values in WR\,135 (197~\AA\ and 185~\AA\ respectively). 

We take relative weakness of the C\,{\sc iii} and C\,{\sc iv} lines as evidence 
for their dilution, in this wavelength region, by the proposed stellar companion. 
Taking the ratios of 5696-\AA\ and 5813-\AA\ lines in WR\,48a and the WC8 standard 
WR\,135, the companion would be $\sim 3 \times$ (1.2 mag.) brighter than the WC8 
star in $V$ and the total luminosity $\sim 4 \times$ (1.5 mag.) brighter. 
If the WC8 star had $M_V = -3.7$ (cf. $M_v = -3.7\pm0.5$ for WC8 stars in clusters 
and associations, van der Hucht 2001), the companion would have $M_V \simeq -4.9$, 
comparable to that of an O6V star (Martins, Schaerer \& Hillier 2005), and the 
system would have $M_V \simeq -5.2$. At 4 pc, this would be $V_0 = 7.8$ which, 
reddening by $A_V = 9$, would give $V=16.8$, close to the $V=17.1$ observed by 
Baume et al. and potentially resolving the discrepancy between the spectroscopic 
distance to WR\,48a and that of the clusters. Fitting the observed $V$ magnitude 
and retaining the adopted distance and reddening would give O5V for the companion 
and $M_V = -4.0$ for the WC8 star, still within the range for its spectral type.  
The total luminosity of the WR\,48a system would be $\sim 4\times 10^5L_\odot$, 
which matches that found by Clark \& Porter (2004).

In an attempt to classify the companion, we re-observed WR\,48a in the blue 
using the ESO Multi-mode Instrument (EMMI) on the 3.6-m New Technology Telescope 
(NTT) at La Silla on 2001 June 17-19 to look for absorption lines. 
The spectrum in the bottom panel of Fig.\,\ref{FOpt} 
is blocked down to 3-\AA\ bins to improve the S/N ratio. The C\,{\sc iii} 
4649-\AA\ triplet is present about 50\% stronger than in the low-resolution 
spectrum but we do not confirm the C\,{\sc iii} 4325-\AA\ line, nor do we see 
any evidence for emission at the position of C\,{\sc iv} 4441\AA\ usually seen 
more strongly in WC8 and still present in WC9 spectra. Evidently, the continuum 
from the companion is dominant in this wavelength region and it is disappointing 
not to observe absorption lines from it, apart from a possible weak line at the 
position of B$\gamma$, which may owe its weakness to partial infill by emission. 

The formal classification of the WC star is clearly WC8 from the ratio of 
the C\,{\sc iii} 5696-\AA\ and C\,{\sc iv} 5813-\AA\ features but it appears 
to be atypical in terms of He\,{\sc i} strength and wind velocity. Without the 
absorption lines, the companion cannot be classified, but estimates of its 
luminosity adopting a distance of 4~kpc and $A_V = 9$ suggests an O5Ve star 
and gives $M_V = -4.0$ for the WC8 star.

\subsection{Infrared spectroscopy}

\begin{figure}
\centering
\includegraphics[width=9cm]{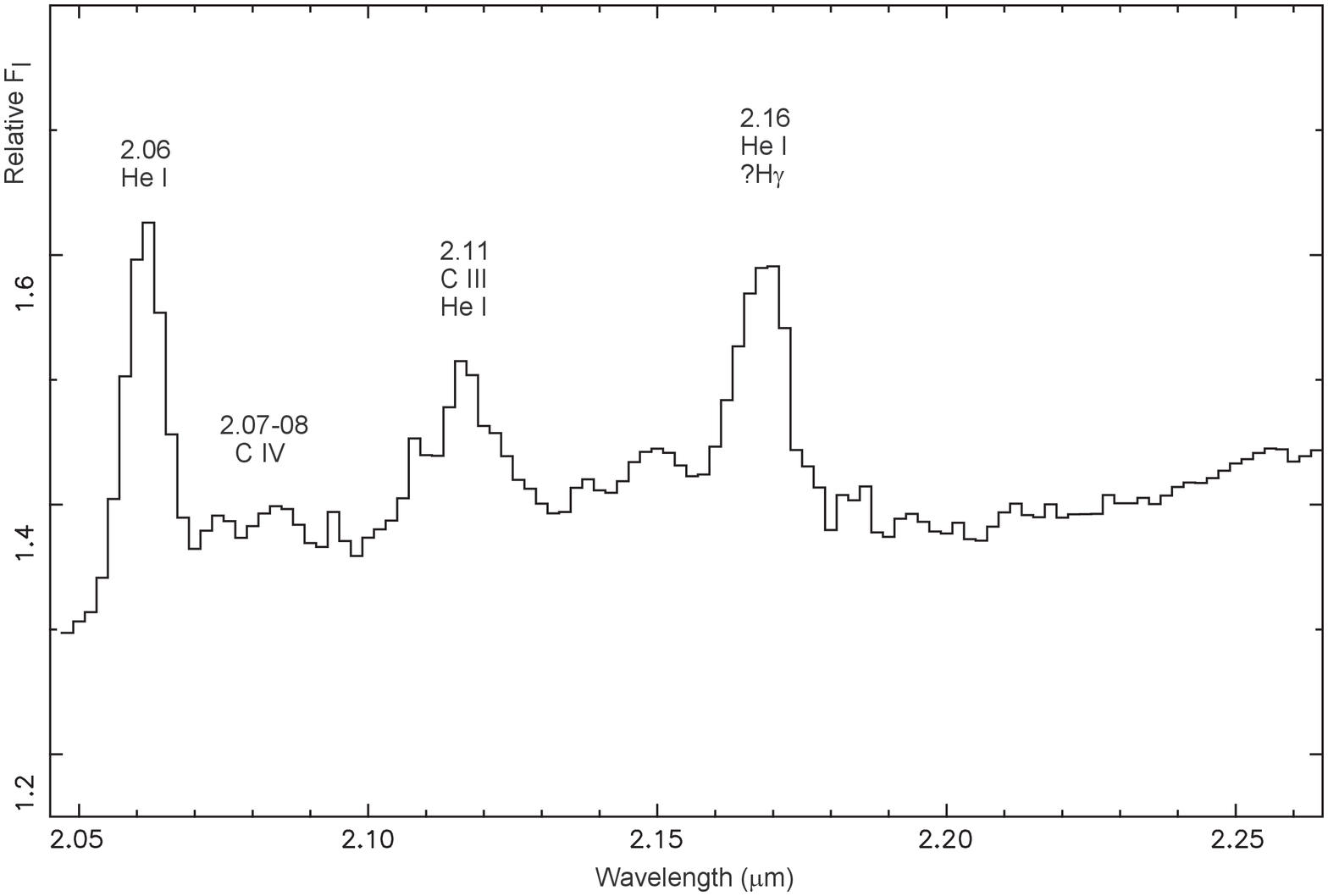}
\caption{Two-micron spectrum of WR\,48a observed with IRIS and formed from 
the 5th and 6th order HK echelle spectra. The spectrum of WR\,48a at this 
wavelength is dominated by the dust emission, most of which is suppressed 
in the figure to show up the lines.}
\label{FIRISK}
\end{figure}

The infrared spectrum of WR\,48a is dominated by dust emission, but it is 
still possible to see the line spectrum on top of this in the near IR. 
Low-resolution NIR spectra were observed for us using the InfraRed Imaging 
Spectrograph (IRIS) on the AAT on 1995 August 15.
The $HK$ echelle gave a spectral resolution of $R = 400$. The spectrum 
in the 2-micron region is shown in Fig.\,\ref{FIRISK}. Comparison with the 
spectra of WC8 and WC9 stars discussed by Eenens, Williams \& Wade (1991) 
and Figer, McLean \& Najarro (1997), particularly the relative strengths 
of the C\,{\sc iv} triplet at 2.07-08 $\mu$m and C\,{\sc iii}/He\,{\sc i} 
blend at 2.11 $\mu$m shows that WR\,48a is a closer match to the WC9 spectra 
than the WC8 star WR\,135. Stronger than either of these features in WR\,48a 
is the He\,{\sc i} 2.06-$\mu$m line, also strong in the WC9 stars WR\,121 
and WR\,119 (cf. Figer et al.) but not WR\,88 (cf. Eenens et al.). 
Of particular interest is the strength of the 2.16-$\mu$m He\,{\sc i} 
feature, which is stronger then the other lines in any of the WC8 or WC9 
spectra in Eenens et al. and Figer et al. We suggest one possible reason 
for this is a contribution from Br $\gamma$ from the companion to the WR star.

\section{Evolution of the dust emission from WR\,48a}
\subsection{Infrared light curves}
\label{SCurves}

\begin{figure*}
\centering
\includegraphics[width=13cm]{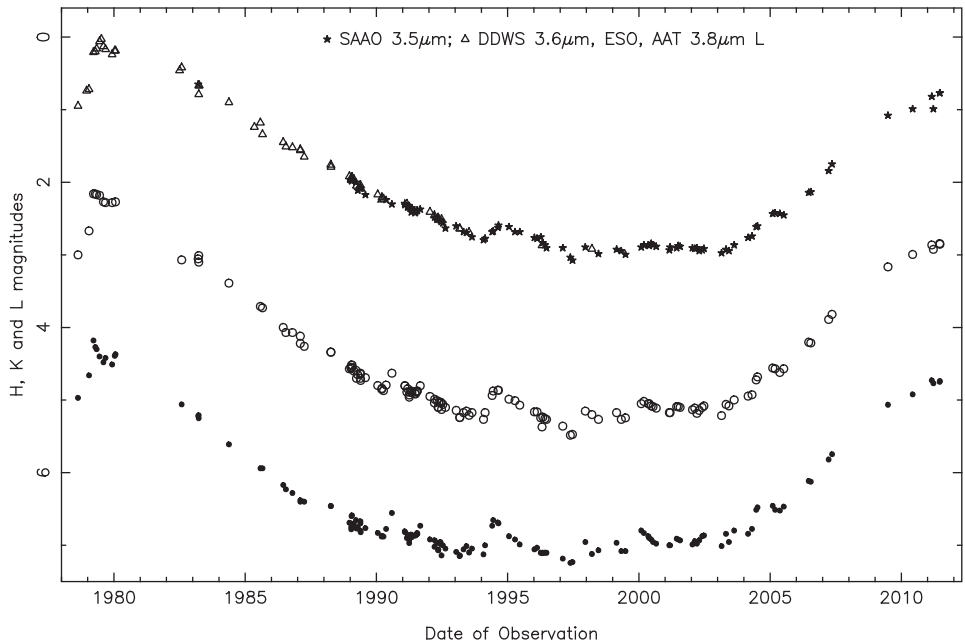}
\caption{Synoptic $H$, $K$ and $L$ light curves of WR\,48a from SAAO data 
corrected for contamination from CPD -62$\degr$ 3058 and the ESO and AAT data 
adjusted as described in the text. Given the long time-scales, we use 
civil dates for discussion of the dust emission. The $L$ magnitudes from 
different sources are plotted with different symbols and show concordance 
where they overlap.}
\label{FHKL}
\end{figure*}

The light curves in $H$, $K$ and $L$ from the homogenized NIR data are shown in 
Fig.\,\ref{FHKL}. We see that the flux faded until about 1997, after which it grew 
very slowly until about 2003, and then began rising again. In $K$ it reached the 
level of DDWS's first (1978) observation in 2010 and continued brightening, 
albeit more slowly. 

Superimposed on the long-term variation are secondary episodes in 1990, 1994--5, 
1999--2000, 2004 and perhaps 1997--8 and 2010--11. The best defined is that in 1994--95. 
From 1994.1, the flux rose to maximum in 1994.4 ($J$ and $H$) to 1994.6 ($K$ and $L$) 
and then gradually faded through 1995 and 1996. The rates of fading are strongly 
wavelength dependent ($\simeq$ 0.38 mag.y$^{-1}$ in $J$, 0.34 mag.y$^{-1}$ in $H$, 
0.22 mag.y$^{-1}$ in $K$ and 0.15 mag.y$^{-1}$ in $L$) and the evolution of the IR 
colour (Section \ref{SCMD}) is consistent with the formation and cooling of extra dust. 
Within the coverage of the observations, the other episodes have similar shapes: a 
steep rise for about 0.4 yr followed by a slower decline. Simple modelling of the 
SED of the epsiode is discussed in Section \ref{SModels}.
To determine the timing of other episodes, we shifted the $H$ and $K$ light curves 
defining the 1994 mini outburst to those near the other events and used the best 
fits to measure the shifts to the other events, deriving --3.95 yr, +3.3 yr, +5.4 yr 
and +10.2 yr, giving intervals between the mini epsiodes of 3.95 yr, 3.3 yr, 2.1 yr 
and 4.8 yr. There is no suggestion of periodicity in these data. 
The $J$, $H$ and $L$ photometry of DDWS suggests that there was another in 1980, but 
this is not evident in their $K$ photometry. It is possible that there 
were mini episodes between then and 1990, but the data are too sparse to be certain.
As discussed in Section \ref{SCMD}, the trajectory in the NIR colour-magnitude 
diagram points to more secondary outbursts, including a possible one in 2010--11. 
These irregularities suggest that the IR light curves will not repeat exactly and 
will make it hard to determine a photometric period without several more years 
observations but the present data suggests a period a little longer than 32 years 
and we will discuss WR\,48a in the framework of its being a long-period CWB.

\begin{figure}
\centering
\includegraphics[width=7cm]{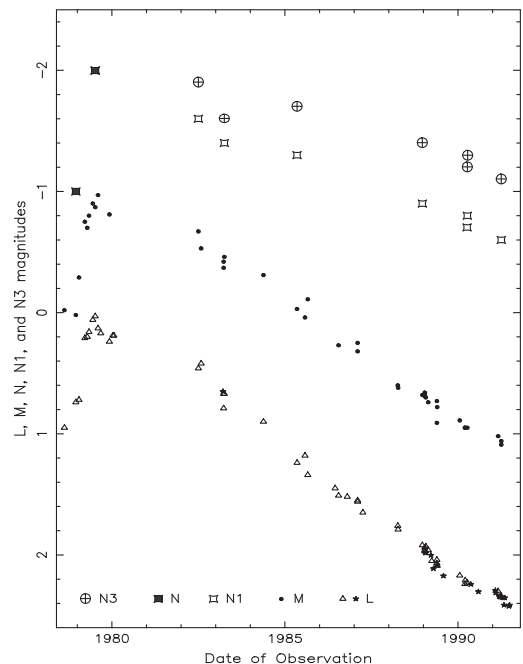}            
\caption{Mid-IR light curves of WR\,48a during 1979 maximum and subsequent fading.
}
\label{LMNN}
\end{figure}

We tracked the fading of the dust emission at longer wavelengths and show light 
curves in $N3$ (12.8$\mu$m), $N1$ (8.4$\mu$m), $M$ and $L$, together with $N$ from 
DDWS in Fig.\,\ref{LMNN}. As can be seen from Table \ref{TESOb}, the $N2$ and $Q$ 
magnitudes follow the same trend ($N2$ is always fainter than $N1$ and $N3$ owing 
to the coverage by this band of the strong silicate absorption feature) but are 
omitted from Fig.\,\ref{LMNN} for clarity. 
Between 1983 and 1991, the rates of fading are strongly wavelength dependent: 
0.07 mag~yr$^{-1}$ in $N3$, 0.08 mag~yr$^{-1}$ in $N1$, 0.16 mag~yr$^{-1}$ in $M$ 
and 0.17 mag~yr$^{-1}$ in $L$, showing continued cooling of the dust.  
Between the 1979 maximum and our first mid-IR photometry the fading appears to 
be slower, and at 20 microns our $Q0 = -1.7\pm0.2$ is comparable to $O = -1.652\pm0.2$ 
observed by DDWS, reminiscent of the behaviour of WR\,140, where the long-wavelength 
fluxes brightened a year after maximum (Williams et al. 2009b).

\subsection{Synopsis of dust evolution from the infrared colour-magnitude diagram}
\label{SCMD}
\begin{figure}
\centering
\includegraphics[width=8.5cm]{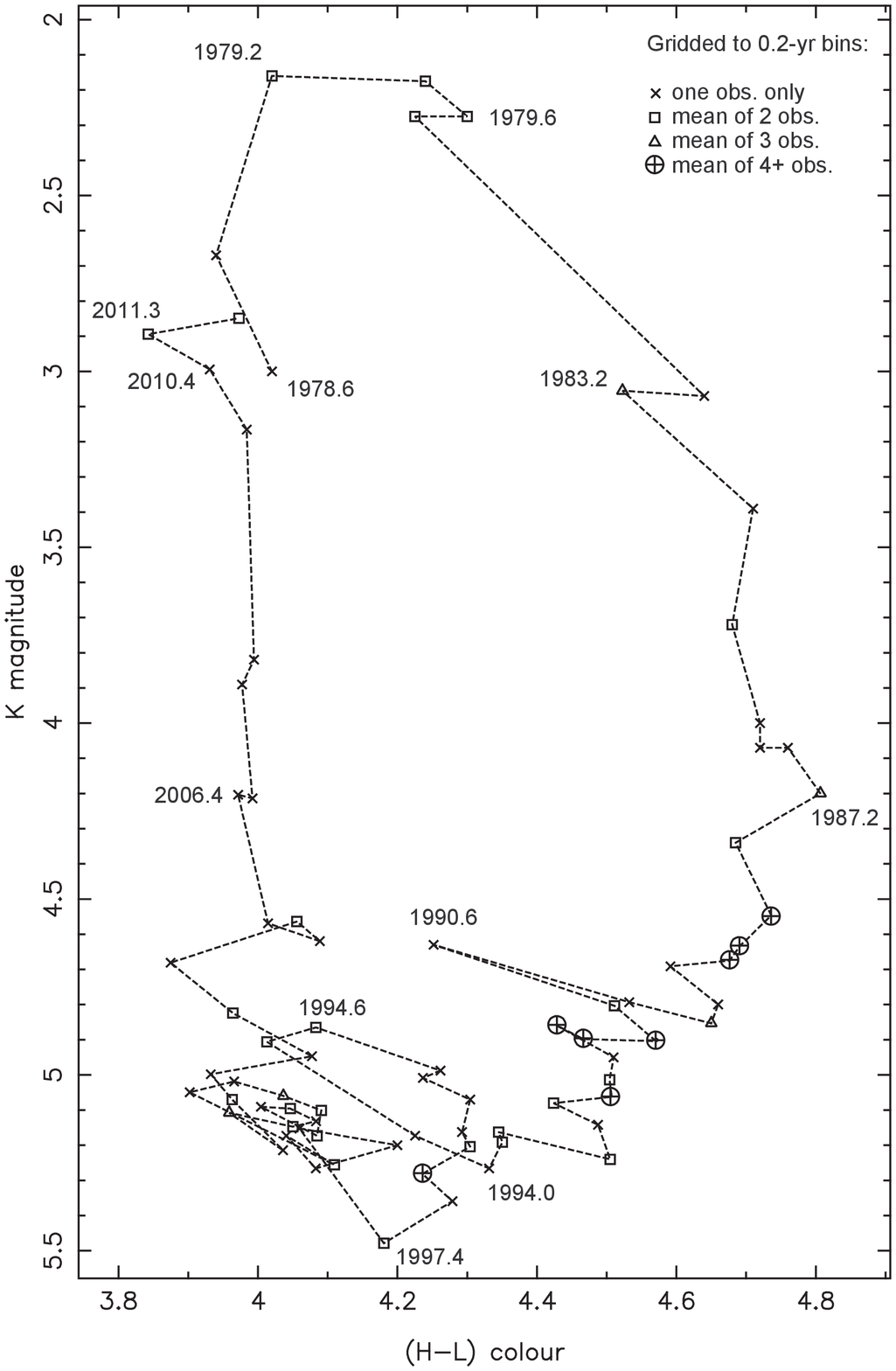}
\caption{Trajectory of WR\,48a in the ($H-L$), $K$ colour-magnitude diagram from 
the first observation by DDWS in 1978 to 2011.5. The data are gridded to 0.2-yr 
bins and the numbers of observations contributing to 
each point are marked with different symbols.}
\label{CMD}
\end{figure}

There are three principal processes determining the evolution of the dust made by 
a system like WR\,48a and its infrared emission (cf. consideration of the dust 
made by the archetypal dust-making CWB WR\,140 by Williams et al. 2009b).  
These are the nucleation of new grains within the compressed stellar winds, the 
growth of the grains by accretion of carbon ions as they are accelerated through 
the wind by radiation pressure, and the cooling of the grains as they move away 
from the stars and the stellar radiation heating them is progressively diluted.  
The nucleation process is not understood, observations indicate that the newest 
grains have temperatures in the region of 1200~K (e.g. WHT, Williams et al. 2009b).
The growth of the grains can also cause them to cool as they become relatively 
more efficient coolants than small grains. Their velocity from the stars is 
initially that inherited from the compressed wind, and is increased by radiation 
pressure on the grains until they reach a drift velocity relative to the wind 
when the radiative acceleration is balanced by supersonic drag.

Compared with the case of WR\,140, we know less about the WR\,48a system, e.g. 
velocity of the dust and its distance from the stars and stellar luminosity, 
and modelling is premature, but we can at least qualitatively identify these 
processes from the SEDs of WR\,48a and its IR colour and magnitude. 

\begin{figure}
\centering
\includegraphics[width=8.5cm]{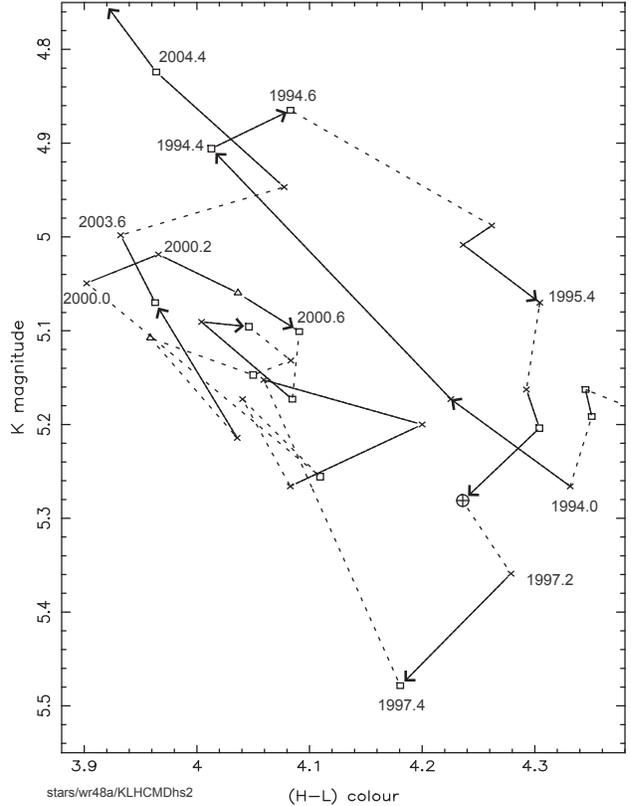}
\caption{Detailed view of WR\,48a in the ($H-L$), $K$ colour-magnitude diagram 
in 1993--2004 showing trajectories around the mini-outbursts in 1994--5, 
1999--2000 and, possibly, 1997--8. The symbols are as in Fig.\,\ref{CMD}, and 
are joined by solid lines when the time bins are consecutive and by broken lines 
when the gaps are greater than 0.2 yr.}
\label{CMDhs}
\end{figure}

We formed a colour-magnitude diagram from our corrected and adjusted photometry 
to track the evolution of the dust. For the amount of hot dust, we chose $K$ 
magnitudes, and straddled this with ($H-L$) as the best indicator of the 
temperature of the newly formed hot dust. 
Sometimes WR\,48a was observed several times in a short interval, so we have 
gridded the data to avoid confusion in the diagram and also increase the accuracy 
of the points where they come from averages of two or more observations. From 
experimentation, we found a grid size of 0.2 yr gave a good balance preserving 
information and providing clarity. The CMD is shown in Fig.\,\ref{CMD}.

Initially, in 1978--1979.2 (DDWS's observations between JD 2443739 and 2443954), 
we observe the flux brightening while the colour did not change. 
We attribute this to the formation of new dust at a higher rate than before. 
The constant colour implies a constant temperature, that at which the dust 
was condensing, and its persistence shows that the SED was dominated by emission 
from the hottest, newest dust, -- i.e. that the nucleation rate exceeded the 
rate at which the dust was carried away in the wind. 
During 1979, the dust emission cooled while the $K$ flux barely changed. 
This can also be seen in DDWS's light curves: after JD 2443954, the $L$ and $M$ 
fluxes grew while $K$ and $H$ fell slightly.  
(DDWS consider their $J$ magnitudes might be affected by emission-line flares.)
We interpret this phase as the continued growth of the newly formed grains 
while the nucleation process itself had slowed; if the grains did not grow, 
then the flux at all the wavelengths would {\em start falling  at the same time} 
as the grains were carried away from the star and cooled as the stellar 
radiation heating them was progressively diluted. 
After 1979 (DDWS's photometry on JD 2444215), fluxes at all wavelengths faded 
and the colour became redder owing to cooling of the dust as the condensation rate 
slowed. The CMD shows continued fading, with ($H-L$) reaching a maximum at 1987.2.
The fall in average dust temperature implied by the reddening in ($H-L$) is 
interpreted as indicating that the dust being continually carried away from the 
star was no longer being replenished by the condensation of new grains at the 
same rate as before. 
After 1987, the ($H-L$) gets bluer again while the $K$ flux continues to fade. 
We interpret this as an indication that the dust formed in the 1979 maximum had 
now cooled sufficiently that its SED was no longer dominating the $L$ band emission.
At longer wavelengths, continued cooling after 1987 can be seen from the the 
mid-IR colours (Table \ref{TESOb}): e.g. ($L^{\prime}-N1$), which had increased 
from 1.6 in 1982 to 2.4 in 1988 (JD 2447519), continued increasing to 2.6 in 1991 
(JD 2448345,7). Some of the SEDs are compared in Fig.\,\ref{Cooling} below.

Sharp blueward movement in ($H-L$) accompanied by rises in $K$ were observed in 
1990 and 1994, the times of the mini episodes seen in the light curves. 
These changes are interpreted as the consequence of additional episodes of dust 
formation: the bluer ($H-L$) being a consequence of the addition of hot dust at 
the condensation temperature, while the total NIR emission rises. 
We have sufficiently frequent observations to track the evolution of the dust 
emission in the 1994--95 event. This is shown in more detail in Fig.\,\ref{CMDhs}. 
After the minimum in ($H-L$), the emission gets redder while $K$ continues rising 
(1994.4--1994.6). This implies that nucleation of grains has ceased but that the 
mass of dust continues to rise as the grains grow by accretion. After 1994.6, 
the emission fades and the dust continues cooling as it is carried away in the 
wind. Part of a similar clockwise trajectory can be seen in 2000.0--2000.6, and 
smaller bits of the others can be discerned. 

After the $K$ minimum in 1997.4, and mini eruptions, the ($H-L$) is increasingly 
influenced by newly formed hot dust and remains blue as the $K$ brightens, with 
a superimposed secondary outburst in 2010.4--2011.3. Generally, however, the 
$K$, ($H-L$) slope in the CMD of the major rise to maximum differs from those 
of the mini episodes in that the brightening $K$ due to increased dust is not 
accompanied by further blueing of ($H-L$): the blue limit must be set by the 
dust condensation temperature. 

\subsection{Spectral Energy Distributions (SEDs)}
\label{SSED}

\begin{figure}
\centering
\includegraphics[width=8cm]{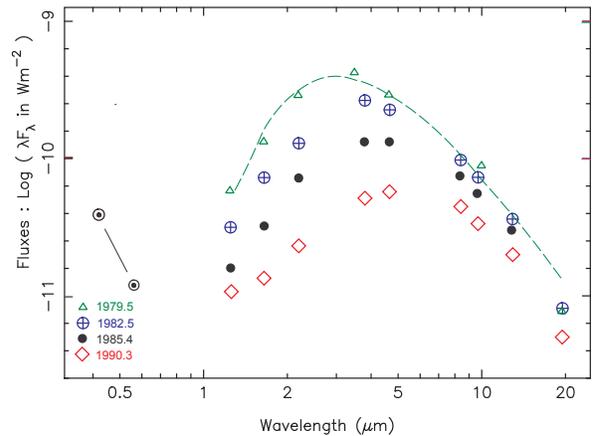}            
\caption{SEDs of WR\,48a as it faded after maximum in 1979: $\triangle$ photometry 
by DDWS in 1979.5; $\oplus$ our mid-IR photometry on JD 2445150 combined with 
DDWS's NIR photometry of 33 days later; $\bullet$ our photometry in mid-1985; and
$\diamond$ our photometry in 1990.3. The optical data come from $B$ and $V$ 
observed in 2006 by Baume et al. (2009), see text. All the photometry was de-reddened 
by $A_V = 9$. The dashed line through the 1979 
points is a 1210-K Planck function for reference.}
\label{Cooling}
\end{figure}

A set of SEDs illustrating the cooling and fading of the dust between maximum and 
our photometry in early 1990 (before the mini dust-formation episode described 
above) derived from the photometry is given in Fig.\,\ref{Cooling}. Following 
the discussion in Section 1, we de-reddened the photometry with a standard 
interstellar law for $A_V = 9.0$.
The dashed line through the fluxes at maximum is a 1210-K Planck function for 
reference -- but it may be a good description of the dust SED at this phase if 
the emission is optically thick. The IR luminosity at maximum 
($5.6\times10^{-11}$Wm$^{-2}$, about $2.8 \times 10^5 L_{\odot}$ at the adopted 
distance of 4~kpc) is about 70 per cent of the total stellar luminosity, 
$\sim 4\times10^5 L_{\odot}$ for the WC8 star plus its proposed O companion. 
This is the fraction of the stellar UV-optical flux that is absorbed and 
re-radiated by the dust, and would give the UV-optical depth in the dust if it 
were uniformly distributed around the stars; but we know from the images that 
the dust is not symmeterical, so some sightlines will have greater optical depth.

A Planckian SED in the $\log\lambda F_{\lambda}$ {\em vs.} $\log\lambda$ plane is 
isomorphic with temperature and flux level, so the broader SEDs seen as the dust 
cools in 1979--90 indicate that we are observing dust having a greater range of 
temperature. 

We can extrapolate the 1979.5 SED shortwards to 1 micron to investigate the 
possible dilution of the 10830-\AA\ and 9710-\AA\ emission lines in DDWS's 
1980 spectrum by dust emission. This extrapolation, reddened by $A_V = 9$, 
gives a dust-emission magnitude at 1 micron of 10.4. 
This is significantly brighter than the 1-micron magnitude ($\sim$ 12.1) 
extrapolated from the 2006 $BVIc$ photometry by Baume et al. (2009) and 
would provide the additional wavelength-dependent continuum emission implied 
by the dilution of the longest wavelength lines in the 1980 spectrum. At 
shorter wavelengths, this is not a problem: the reddened dust-emission $Ic$ 
is 14.8, significantly fainter than the $Ic = 13.25$ observed by Baume et al.

\begin{figure}
\centering
\includegraphics[width=8cm]{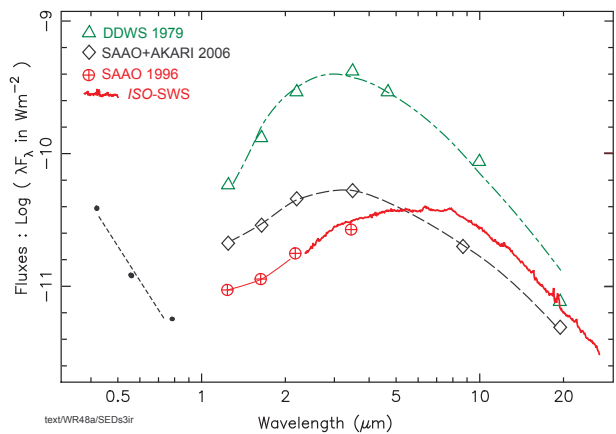}            
\caption{SEDs of WR\,48a at three phases: $\triangle$ the IR maximum 1979.5 observed 
by DDWS; $\oplus$ close to IR minimum from our $JHKL$ photometry in mid-1996 soon 
after the {\em ISO}-SWS spectrum from Paper~I, solid line; and $\diamond$ while 
the IR flux was rising from our $JHKL$ photometry in mid-2006 with 9-$\mu$m and 
18-$\mu$m fluxes observed by {\em AKARI} in 2006--07. 
The optical flux points come from $BVIc$ photometry by Baume et al. in 2006, with 
$B$ and $V$ supported by HST WFPC2 $439W$ and $556W$ photometry in 1996, see 
text. All the photometry was de-reddened by $A_V = 9$; the {\em ISO} spectrum was 
der-reddened by nulling the 9.7-$\mu$m silicate feature.}
\label{SEDs3}
\end{figure}

Even at minimum, we can see from the spacing of light curves that there was still 
significant dust emission by WR\,48a. We examine the SED in 1996, near IR minimum 
by combining the SAAO $JHKL$ photometry observed on JD 2450127, contemporaneously 
with the {\em ISO}-SWS spectrum presented in Paper~I. Convolution of the SWS 
spectrum with the $L$ filter passband recovers the observed $L$ magnitude. 
The SED (Fig.\,\ref{SEDs3}) showing strong dust emission, and the absence of further 
fading after 1997 to the stellar wind level, demonstrate continuing dust formation, 
which replenishes that carried away in the stellar wind. 
In this respect, WR\,48a differs from the episodic dust makers like WR\,140, and 
resembles the variable dust-maker WR\,98a discussed in Paper I.
 
To estimate the SED during the rising branch of the light curve, we retrieved mid-IR 
fluxes for WR\,48a observed (as AKARI-IRC-V1 J1312398-624255) in the {\em AKARI} 
IRC Survey (Ishihara et al. 2009) from the ISAS/JAXA Archive. These data come from 
six scans made between 2006 May and 2007 August and we combined them with $JHKL$ 
observed in 2006 (JD 2453906 and 2453935). Comparison of the {\em AKARI} 9-$\mu$m 
and 18-$\mu$m fluxes with the {\em ISO}-SWS SED (Fig.\,\ref{SEDs3}) suggests that 
the mid-IR flux from WR\,48a barely changed over the decade between the two 
missions, although we note that the breadth of the $S9W$ and $L18W$ filters used 
for the IRC survey and their coverage of the `silicate' bands requires 
care in the conversion of in-band fluxes to monochromatic fluxes for a SED and 
the observed and extinction-corrected {\em ISO}-SWS spectra need to be convolved 
with the {\em AKARI} IRC optics to examine this further. 
For the present, we note that while the mid-IR flux barely changed, the near-IR 
has increased significantly between 1996 and 2006, indicating the formation of 
new hot dust. 

The optical data in Figs \ref{Cooling} and \ref{SEDs3} ($B$, $V$ and $Ic$) come 
from the photometry in 2006 by Baume et al. (2009). 
Their $V = 17.13$ agrees well with the HST WFPC2 magnitude $F555W = 17.08$ observed 
on 1996 July 4\footnote{Date is from the HST archive.} by Wallace et al. (2003), 
while the difference between their $B = 19.37$ and Wallace et al.'s $F439W = 19.79$ 
is readily understood in terms of the longer wavelength cutoff of the $B$ filter 
and the red colours of WR\,48a, which lie outside the range of colours for which 
the transformations between WFPC2 and $UBV$ colours given by Holtzman et al. 
(1995) apply. Hence, the $B$ and $V$ are a good measure of the optical flux from 
WR\,48a at the times of our SEDs in both 2006 and 1996 at least. 
We note that the $Ic = 13.25$ measured by Baume et al. is significantly fainter 
than the $i$ = 12.50 and 12.58 observed in the DENIS survey (Epchtein et al. 1999) 
in 1997 and 1998 respectively. 
This cannot be emission by the hot dust as we can see (Fig.\,\ref{SEDs3}) that 
there was more hot dust emission in 2006 than in 1997, and further observations in 
this wavelength region are needed to look for variability.

\subsection{Modelling the SED of the `mini' eruption}
\label{SModels}

At its simplest, the IR SED of dust emission is given by the convolution 
of the grain emissivity, $\kappa_{\lambda}$, and the Planck function for 
the dust temperature, $T_{\rmn{g}}$: 
\[
4 \pi d^2 F_{\lambda} = m_{\rmn{d}} \kappa_{\lambda} B(\lambda, T_{\rmn{g}})
\]
\noindent where $d$ is our distance to the dust, $m_{\rmn{d}}$ is the dust 
mass and the emission is optically thin in the IR. Given $\kappa_{\lambda}$, 
the temperature of the grains can be found by fitting the SED over a suitable 
wavelength range, followed by the dust mass from the overall fit and distance, $d$. 
We calculated $\kappa_{\lambda}$ from the optical properties of the `ACAR' 
amorphous carbon grains prepared in an inert atmosphere in the laboratory 
by Colangeli et al. (1995) and tabulated by Zubko et al. (1996). 
Model SEDs calculated using the `ACAR' data gave a better fit to broad 
features in the 5.5--6.5 $\umu$m {\em ISO} spectra of dust-making WC8--10 
stars (Williams, van der Hucht \& Morris, 1998) than those based on the 
`ACH2' grains produced by Colangeli et al. in a hydrogen atmosphere --
as one might expect for grains formed in hydrogen-poor WC stellar winds. 
A comparison of several laboratory analogues of cosmic amorphous carbon grains 
by Andersen, Lodl \& H\"ofner (1999) shows that their absorption efficiencies 
have fairly similar wavelength dependencies, $\kappa \propto \lambda^{-1.1}$,  
in the IR, but absolute values ranging by a factor of $\sim$ 5. Consequently, 
the grain temperatures derived from fitting IR SEDs are relatively insensitive 
to the choice of grain sample, but not the masses of the dust clouds.
Usually, circumstellar grains are not all at the same temperature, but have a 
range of temperature depending on their distances, $r_*$, from the star(s) 
heating them, and also on their size. For small `ACAR' grains and in the absence 
of significant UV-visual optical depth, the temperatures of grains in thermal 
equilibrium with the stellar UV-visual radiation field falls off with distance 
as $T_{\rmn{g}} \propto r_*^{-0.38}$ (Williams et al. 2009). This dependence 
is also relatively insensitive to grain sample, and was found to apply on 
average even in the full radiative transfer modelling of the thick dust spiral 
of the classical persistent dust-making WC9 star WR\,104 by Harries et al. (2004).

Similar techniques will be required for modelling of the dust made by WR\,48a 
owing to its optical depth. This will require knowledge of the dust geometry, and 
is beyond the scope of the present paper; for the present, we consider simple 
optically thin, isothermal modelling of the 1994--5 `mini' eruption. We model the 
SEDs at two epochs: the maximum in $J$ and $H$ at 1994.45 (data from JD 2449519) 
and that in $K$ and $L$ in 1994.64 (averaging data of JD 2449583 and 2449590). 
This emission is in excess of the continuous `baseline' dust emission 
by WR\,48a at the time, and we formed a `baseline' SED interpolating between 
the light curves before and after the `mini' dust-formation episode. 
The additional IR luminosity in the 1994 sub-peak at maximum (i.e. subtracting 
the baseline emission) was $\sim$ 10$^{-11}$ Wm$^{-2}$, equivalent to 
$\sim 5 \times 10^3 L_{\odot}$ at the adopted distance of 4~kpc, is only one 
per cent of the total stellar luminosity, so we treat its emission as being 
optically thin -- although it may suffer extinction in the cooler dust in our 
sightline. Also, if the dust is newly formed, the spread in $r_*$ and grain 
size may be small enough for the dust to be approximately isothermal, and we 
can usefully derive an average temperature $\langle T_{\rmn{g}} \rangle$ from 
the observed SED. 

The results of fitting these differential SEDs are given in Table \ref{Mods}, 
in which the uncertainties were estimated from those in the photometry 
(0.02 mag. in $JHK$ and 0.03 mag. in $L$). The two epochs roughly separate 
three phases in the evolution of the dust: (a) nucleation of new grains near 
the stars until 1994.45, (b) growth of the grains as they moved away from the 
stars causing the dust mass to increase between then and 1994.64 and (c) the 
continued fading of the dust emission. The grain temperature from the 1994.45 fit 
($\langle T_{\rmn{g}} \rangle = 1220\pm$90~K) is close 
to that at which the grains nucleate; the relatively large uncertainty comes 
from fitting the small difference between the two SEDs, and there must be 
some range in temperature as the first grains to form will have moved further 
from the stars. There was no emission from the `mini' episode in the 1994.06 
photometry, so we determine an average dust formation rate, presumably a 
combination of nucleation and growth of the oldest grains, for this phase.  
The lower $\langle T_g \rangle$ two months later results primarily from the 
movement of the mini dust cloud away from the stars.  There may be a component 
from more efficient cooling by the largest grains, but we do not have the data 
(e.g. velocity of the dust grains) to separate the contributions. The formation 
rate in this second phase is about four times that in the first and must be 
almost entirely due to the growth of the existing grains as the $J$ and $H$ 
fluxes were falling at the time, implying that nucleation had ceased. The 
three-fold increase in dust mass implies a 40\% increase in average grain 
radius in the two months of the second phase. 

The average dust-formation rate, $1.4\pm0.3 \times 10^{-7} M_{\odot}$ yr$^{-1}$, 
seems high for a minor dust-formation episode compared with the likely mass-loss 
rate of WR\,48a ($\sim  10^{-5} M_{\odot}$ yr$^{-1}$) and abundance of carbon in 
its wind, but we recall the range in the scales of $\kappa_{\lambda}$ for 
different laboratory grain samples resulting in a similar range of derived 
dust mass. 

\begin{table}
\caption{Isothermal model fits to $JHKL$ SEDs during the 1994 mini dust-formation episode.}
\label{Mods}  
\centering            
\begin{tabular}{lrcc}
\hline
Date    & $\langle T_d \rangle$ &  $M_d$                   & formation rate         \\
        &          (K)          &  ($M_{\odot}$)           & ($M_{\odot}$ yr$^{-1}$) \\
\hline
1994.45 &     1220$\pm$90 &     $1.4\pm0.7 \times 10^{-8}$ & $3.6\pm1.8 \times 10^{-8}$  \\    
1994.64 &     1050$\pm$20 &     $4.1\pm0.8 \times 10^{-8}$ & $1.4\pm0.3 \times 10^{-7}$  \\  
\hline
\end{tabular}
\end{table}

\section{Discussion} 
\label{SDisc} 

The long-term infrared photometry of WR\,48a shows complex behaviour different 
from that of any other WR star: a slow, large-amplitude variation on which are 
superimposed briefer variations of lower amplitude. In all cases in which they 
can be determined, the trajectories in the ($H-L$), $K$ colour-magnitude diagram 
are loops (clockwise in Figs \ref{CMD} and \ref{CMDhs}) indicative of the formation 
of new, hot dust, followed by its cooling. The recurrence time-scale for the 
slow variation is 32+ yrs; although the 2010--11 NIR photometry reached the 
levels of the earliest (1978--79) pre-maximum photometry by DDWS, 
the interval cannot be translated into a period owing to the superimposed 
short-term variations, and further observations capturing at least the maximum 
and early decline are needed for a period\footnote{Strictly, observations over 
another whole cycle, but these would take the first author to his centenary and 
are not planned}.
Even without a period, there is enough evidence to consider WR\,48a as a CWB 
periodic dust-maker and discuss it in that context.

The total dust luminosity at maximum (1979) is 13$\times$ that in 1996, near 
minimum, implying a variable rate of dust formation assuming a constant wind 
dispersing the dust. The rate of dust formation is better estimated from the 
short wavelength flux sensitive to the hottest, newly formed dust and the range 
in $H$ between maximum and minimum (adjusted for the stellar contribution) 
gives a factor $\simeq 16$. The fraction of the WC8 wind going into the WCR 
is determined by the angular size of the WCR, which depends on the wind-momentum 
ratio $\eta = (\dot{M}v_\infty)_{OB}/(\dot{M}v_\infty)_{WC}$,  
which does not change around the orbit if the winds collide at their terminal 
velocities, but the pre-shock wind density at the WCR does change with the 
stellar separation if the orbit is elliptical. If maximum and minimum dust 
formation coincide with periastron and apastron passages in an elliptical 
orbit, and if the dust formation rate is proportional to the pre-shock wind 
density at the WCR, the stellar separation varies by a factor of $\simeq 4$ 
and the orbital eccentricity would be $e \simeq 0.6$, with periastron passages 
in 1979 and some time after mid-2011.

The mini eruptions appear not to be periodic, arguing against the suggestion 
in Paper I of a third stellar component in WR\,48a, and other large-scale 
overdensity structures which are intercepted by the WCR as it moves through 
the winds are implied. 
The only other system having mini outbursts superimposed on its long-term 
dust-formation variation is WR\,137. This is one of the few WR stars to have 
significant continuum polarization and wind asymmetry (Harries, Hiller \& 
Howarth 1998). Harries, Babler \& Fox (2000) pointed out the similarity of 
WR\,137 to WR\,6 and WR\,134, which display periodic variability in their 
emission-line morphology possibly caused by the rotation of large-scale 
azimuthal structures within their flattened winds. Polarization observations 
of WR\,48a are needed to to see if this is the case here. Also, the H$\alpha$ 
emission associated with the companion suggests that this might have a 
flattened wind and large-scale structures.

The persistence of dust formation even at IR minimum classifies WR\,48a as a 
variable dust maker like WR\,98a, which has a period near 564 d. determined 
from both the rotation of its dust pinwheel (Monnier et al. 1999) and NIR 
photometry (Paper~I). The recurrence time-scale of WR\,48a much longer than 
this, and the periods of the episodic dust makers, of which the longest is 
13.05 yrs for WR\,137, both from IR photometry (Williams et al. 2001) and the 
RV orbit (L\`efevre et al. 2005), and even the orbital period, 24.8 yr, of 
the persistent dust-maker WR\,112 inferred from its multi-arc dust cloud 
(Marchenko et al. 2002). The relevance of the long period is 
the implied size of the orbit (for stellar systems of similar mass), with 
implications for the wind-collision process, particularly whether the post-shock 
WC wind is radiative or adiabatic. 

Radiative cooling in the post-shock WC wind is important for dust formation 
(Usov 1991). It was discussed by Stevens, Blondin \& Pollock (1992), who 
introduced the cooling parameter, $\chi$: the ratio of the time-scales for 
cooling of the shocked gas and its escape from the apex of the WCR to flow 
downstream, and given by
\[
\chi = \frac{v^4_8d_{12}} {\dot{M}_{-7}}
\]
\noindent
where $v_8$ is the wind velocity in units of 1000 km~s$^{-1}$, $d_{12}$ is 
the separation of the stars in units of $10^{12}$ cm, and $\dot{M}_{-7}$ 
is the star's mass-loss rate in units of  $10^{-7}$ M$_{\odot}$ yr$^{-1}$. 
Stevens et al. considered a wind with $\chi \geq 1$ to be adiabatic and 
$\chi \ll 1$ to be radiative; recent hydrodynamical simulations by Parkin 
\& Pittard (2008) find a higher threshold, with the compressed wind 
cooling rapidly if $\chi \leq 3$. The WCR in a system having a very 
elliptical orbit, like WR\,140, can vary between adiabatic and radiative 
as the stellar separation varies, becoming radiative close to periastron 
-- the only time that dust forms in that system. 
In the case of WR\,48a we know none of $v_8$, $d_{12}$ and $\dot{M}_{-7}$, 
but can estimate $\chi$ very roughly. If the stars in WR\,48a have masses 
comparable to those in the $\gamma$ Vel system they would have 
$d_{12} \simeq 530$ in a $P = 33$ yr period circular orbit. Adopting a wind 
velocity of 1200 km~s$^{-1}$ (Section \ref{SOpt}) and assuming a mass-loss 
rate of $10^{-5}$ M$_{\odot}$ yr$^{-1}$, we have $\chi \simeq 11$ for the 
WR\,48a WCR. This is significantly above the threshold and indicates an 
adiabatic WCR -- which does not favour dust formation. In an elliptical 
orbit, the separation and hence $\chi$ would be smaller near periastron 
(e.g. $\chi \simeq 4$ for $e = 0.6$); but they would be larger at apastron 
-- and we know from the observations that dust forms throughout the orbit. 
The dust formation by WR\,112 raises a similar problem: for the same 
stellar masses, the stars would have $d_{12} \simeq 390-490$ in an orbit  
having $e = 0.11$ (Marchenko et al. 2002) and $\chi \simeq 8-10$ for the 
same estimates of mass loss and velocity, which are also those used in 
their modelling by Marchenko et al. Higher mass-loss rates would result 
in lower values of $\chi$, as would lower wind velocities, which enter with 
the fourth power. Wind velocities below $\sim 900$ km~s$^{-1}$ would give 
values of $\chi \simeq 3$ and possibly radiatively cooling winds. 
The velocities of both stars need to be measured from high-resolution 
spectra to advance the question of the cooling of the post-shock winds 
and the conditions allowing their dust formation. Velocities are also 
needed to aid interpretation of the dust images of the two stars. 

The recurrence of dust formation, the dilution of the WC emission lines 
implying presence of a luminous companion, the anomalous emission in 
H$\alpha$ (and possible H$\beta$ and Br$\gamma$) supporting this, and 
the strong X-ray emission reported by Zhekov et al. (2011) and Mauerhan et al. 
(2011) all point to WR\,48a being a long-period CWB. The IR photometry needs 
to be extended to determine a period and characterise the `mini' outburst 
and the spectrum needs to be re-observed at higher resolution and S/N to 
elucidate the types of both components and determine the WC wind velocity 
to further the consideration of the conflict between observed dust formation 
and an apparently adiabatic post-shock wind. Knowledge of the wind cooling 
will also be valuable in understanding the X-ray flux from WR\,48a and its 
long term variation. The long period of the system suggests that any non-thermal 
radio emission from the WCR  will have little difficulty in escaping the 
wind (cf. Dougherty \& Williams 2000). Its distance is comparable to 
that of WR\,112 (4~kpc, van der Hucht 2001), which shows variable non-thermal 
radio emission at the level of $\sim$ a few mJy (Chapman et al. 1999, 
Monnier et al. 2002), so we expect the similar flux levels from WR\,48a 
if the stars have similar winds, but with greater amplitude of variation 
given the greater eccentricity and inclination of the WR\,48a system.


\section*{Acknowledgments}

The long-term observing campaign reported here would have been impossible without 
the active support of many individuals and observatories. We would like to thank 
the Directors and Staff at the SAAO, ESO and the AAO for continued access to 
facilities and their support over a long period, and Pik Sin Th\'e, Frank Spaan, 
R. de Jong and D. de Winter for contributing or helping with the observations at ESO. 
It is a pleasure to thank Martin Cohen for helpful discussions. 
This research has made use of the NASA / IPAC Infrared Science Archive, 
which is operated by the Jet Propulsion Laboratory, California Institute 
of Technology, under contract with the National Aeronautics and Space 
Administration; the Vizier database, operated by the CDS, Strasbourg, 
the data archive at the Space Telescope Coordinating Facility (ST-ECF) 
at ESO, and the DARTS archive developed and maintained by C-SODA at ISAS/JAXA.
PMW is grateful to the Institute for Astronomy for hospitality and continued 
access to facilities of the Royal Observatory, Edinburgh. PAW acknowledges 
a grant from the SA National Science Foundation.

\end{document}